\documentclass[fleqn,usenatbib]{mnras}
\pdfoutput=1 
\usepackage{newtxtext,newtxmath}

\usepackage[T1]{fontenc}

\usepackage{graphicx}	
\usepackage{color}

\newcommand{\hmpc}{h^{-1}{\rm Mpc}}
\newcommand{\ihmpc}{\,h{\rm Mpc}^{-1}}
\newcommand{\hgpc}{h^{-1}{\rm Gpc}}
\newcommand{\mbx}{\boldsymbol x}
\newcommand{\mbr}{\boldsymbol r}
\newcommand{\cdf}{{\rm CDF}}

\newcommand{\nn}{{\rm NN}}

\newcommand{\eq}[2]{\begin{align} \label{eq:#1} #2 \end{align}}

\newcommand{\quijote}{\textsc{Quijote}\,}

\newcommand{\redmagic}{\texttt{redMaGiC} }
\newcommand{\bq}{\boldsymbol q}
\newcommand{\bx}{\boldsymbol x}

\newcommand{\bPsi}{\boldsymbol{\Psi}}

\newcommand{\knncdf}{$k\nn$-$\cdf$ }

\title[Modeling kNN distributions with HEFT]{Modeling Nearest Neighbor distributions of biased tracers using Hybrid Effective Field Theory}

\author[Banerjee et. al]{
Arka Banerjee $^{1}$\thanks{E-mail: {\tt arka@fnal.gov}}, Nickolas Kokron $^{2,3,4}$\thanks{E-mail: {\tt kokron@stanford.edu}}
and Tom Abel $^{2,3,4}$\thanks{E-mail: {\tt tabel@stanford.edu}} \\
$^{1}$Fermi National Accelerator Laboratory, Cosmic Physics Center, Batavia, IL 60510, USA \\
$^{2}$Kavli Institute for Particle Astrophysics and Cosmology, Stanford University, 452 Lomita Mall, Stanford, CA 94305, USA \\
$^{3}$Department of Physics, Stanford University, 382 Via Pueblo Mall, Stanford, CA 94305, USA \\
$^{4}$SLAC National Accelerator Laboratory, 2575 Sand Hill Road, Menlo Park, CA  94025, USA
}

\date{Accepted XXX. Received YYY; in original form ZZZ}

\pubyear{2021}
\begin{document}
\label{firstpage}
\pagerange{\pageref{firstpage}--\pageref{lastpage}}
\maketitle

\begin{abstract}
We investigate the application of Hybrid Effective Field Theory (HEFT) --- which combines a Lagrangian bias expansion with subsequent particle dynamics from $N$-body simulations  --- to the modeling of $k$-Nearest Neighbor Cumulative Distribution Functions ($k\nn$-$\cdf$s) of biased tracers of the cosmological matter field. The $k\nn$-$\cdf$s are sensitive to all higher order connected $N$-point functions in the data, but are computationally cheap to compute. We develop the formalism to predict the $k\nn$-$\cdf$s of discrete tracers of a continuous field from the statistics of the continuous field itself. Using this formalism, we demonstrate how $k\nn$-$\cdf$ statistics of a set of biased tracers, such as halos or galaxies, of the cosmological matter field can be modeled given a set of low-redshift HEFT component fields and bias parameter values. These are the same ingredients needed to predict the two-point clustering. For a specific sample of halos, we show that both the two-point clustering \textit{and} the $k\nn$-$\cdf$s can be well-fit on quasi-linear scales ($\gtrsim 20 \hmpc$) by the second-order HEFT formalism with the \textit{same values} of the bias parameters, implying that joint modeling of the two is possible. Finally, using a Fisher matrix analysis, we show that including $k\nn$-$\cdf$ measurements over the range of allowed scales in the HEFT framework can improve the constraints on $\sigma_8$ by roughly a factor of $3$, compared to the case where only two-point measurements are considered. Combining the statistical power of $k\nn$ measurements with the modeling power of HEFT, therefore, represents an exciting prospect for extracting greater information from small-scale cosmological clustering.
\end{abstract}




\section{Introduction}
\label{sec:intro}

Various cosmological surveys, both current such as EBOSS, DES, and KiDS \citep[see][]{2021PhRvD.103h3533A,2021arXiv210513549D,2021A&A...646A.140H} and upcoming ones, including LSST, Euclid, SPHEREx, WFIRST, DESI, and PFS \citep{2019ApJ...873..111I,2011arXiv1110.3193L,2014arXiv1412.4872D,2019BAAS...51c.341D,2016arXiv161100036D,2014PASJ...66R...1T}, are in the process of mapping out the clustering of tracers of structure in the Universe, such as galaxies, in progressively greater detail. This data can be used to answer various fundamental questions, such as the equation of state of Dark Energy, the nature of Dark Matter, and total mass of the Standard Model neutrinos. Much progress has been made on this front in the last few decades, using the clustering of structure on large scales, where linear perturbation theory (in the density contrast) provides an accurate framework. The next challenge is to model clustering on smaller scales --- where linear perturbation theory breaks down --- and thereby extract maximal information from the data provided by these surveys. It should also be noted that once the underlying cosmological matter field is no longer linear, the approximation that it is a Gaussian random field breaks down. This implies that the $2$-point correlation function, which is the complete statistical description of a Gaussian random field, does not characterize the field completely. Therefore, the challenge of extracting useful information on the cosmological parameters of interest from from small-scale clustering is intimately connected to characterizing clustering beyond the $2$-point correlations  \citep{10.1093/mnras/stv2506,2020JCAP...03..040H,2021JCAP...04..029H,2021JCAP...05..030C,2020MNRAS.495.4006U}.

One approach to the goal of harnessing smaller scale information is through the use of $N$-body simulations \citep{10.5555/62815}, which accurately model the late-time clustering of the Cold Dark Matter (CDM) component of the Universe down to small and extremely nonlinear scales (see e.g. \cite{2005ApJS..160...28H,2008CS&D....1a5003H}). However, these simulations do not capture the evolution of the baryonic component of the Universe, and therefore, do not by themselves predict the clustering of visible tracers of structure, like galaxies, that are observed by cosmological surveys. To connect the outputs of the numerical simulations and observations, an additional layer of modeling, i.e. the \textit{tracer--matter} connection is needed, and various such prescriptions exist in the literature (see \cite{wechslertinker} for an extensive review). One of the most popular models, especially to model clustering of tracers on intermediate, quasi-linear scales, is the generic \textit{bias expansion} (see \cite{dejacquesetal} for a review of these models). In particular, the Lagrangian bias formalism \citep{PhysRevD.78.083519} connects the late-time clustering of biased tracers to properties of the initial distribution of the underlying matter field. Traditionally, the Lagrangian bias approach has been coupled with perturbation theory approaches to computing the displacement field connecting the early and late time matter fields. An analytic prescription for displacements results in the well-known Lagrangian Perturbation Theory \citep{Matsubara_2008, PhysRevD.78.083519}. Extending the model to include counter-terms due to small-scale dynamics leads to Lagrangian Effective Field Theory \citep{Porto_2014, Vlah_2015}. Other recent developments include formulations of LPT which partially re-sum contributions while preserving Galilean invariance, leading to Convolution Lagrangian Effective Field Theory \citep{Carlson_2012, 2016JCAP...12..007V}. 

Perturbative approaches have been developed over the years to improve models both tracer--tracer and tracer--matter $2$-point clustering. In particular, Eulerian approaches to modeling $2$-point clustering have recently been successfully applied to the analysis of the Baryon Oscillation Spectroscopic Survey (BOSS)\footnote{http://www.sdss3.org/surveys/boss.php} data \citep{2020JCAP...05..005D,2020Ivanov}. These studies represented the first analyses to use the full shape information for the $2$-point clustering of the BOSS data, down to scales of $k_{\rm max} \simeq 0.25\, h {\rm Mpc}^{-1}$. Despite the tremendous advances in perturbative modeling, it is known that by themselves they cannot be used to extract all of the small-scale information contained in the galaxy density field. Perturbative models are best suited to the scales mentioned above  ($k\sim 0.25 \ihmpc$), which in the next generation of spectroscopic surveys such as DESI will be probed with unparalleled volumes and statistical power. Nevertheless, the amount of potential cosmological information contained within correlations on even smaller scales will potentially be even higher. This motivates models that can accurately predict the statistics of tracers down to scales smaller than what can be achieved purely via perturbation theory.

As a way to combine the generality of Lagrangian bias models with smaller-scale information, \cite{modichenwhite} demonstrated that using the displacement field from $N$-body simulations, which accounts for the full gravitational evolution down to nonlinear scales, can increase the range of scales over which the $2$-point clustering of biased tracers is well-described for the same order of Lagrangian bias expansion. This was found to be especially true for moderately biased tracers at low redshifts, where the model worked well up to $k_{\rm max} \sim 0.6 \ihmpc$, nearly doubling the range of scales from perturbation theory approaches. This approach, Hybrid Effective Field Theory (HEFT), has since been merged with an emulator framework \citep{kokron2021,2021arXiv210112187Z}, and recently applied to data \citep{boryana}. While the existing studies all focused on $2$-point clustering, the model applies the bias expansion at the field level, and then uses $N$-body evolution to compute the advected fields. The late-time nonlinear component fields themselves are natural predictions of this model, rather than any particular statistics over these fields.  Therefore, it is of great interest to assess their ability in capturing more complex summary statistics of tracers than just their 2-point functions. Given a set of bias coefficients in HEFT, the fields should also model higher-order statistics (such as the bispectrum and trispectrum) of the tracer population. This was explored, for a similar model to HEFT, in the context of halo samples in \cite{Schmittfull_2019} and for redshift-space samples of galaxies in \cite{schmittfull2020modeling}. 
Nevertheless, the computational complexity of $N > 2$-point analyses is quite large (however, see \cite{2021arXiv210610278P} for recent progress in this direction), and so we seek an alternative summary statistic that encodes higher order information, to test the applicability of HEFT.

\cite{Banerjee_Abel,Banerjee_Abel_cross} have proposed a set of new summary statistics to capture the auto and joint clustering of various tracer samples in the context of cosmology. These statistics, the $k$-Nearest Neighbor Cumulative Distribution Functions ($k\nn$-$\cdf$s) are constructed by considering the distribution of distances to the $k$-th nearest data point from a volume-spanning set of query points. If the tracers represent a local Poisson sampling of an underlying field, the $k\nn$-$\cdf$s are sensitive to integrals of \textit{all} connected $N$-point functions of the underlying field. For two sets of tracers, and their corresponding underlying fields, \cite{Banerjee_Abel_cross} demonstrated that the joint $k\nn$-$\cdf$s are sensitive to all combinations of $N$-point functions that can be formed from the two fields. This makes these statistics a powerful probe of clustering, and was recently applied to the measurement of the clustering of very massive galaxy clusters \citep{2021arXiv211204502W}. At the same time, these statistics are particularly easy to compute on a given dataset, taking orders of magnitude lower computational resources compared to other statistical measures of clustering beyond the $2$-point function. These attractive features make these statistics a promising candidate to apply to cosmological datasets from surveys, especially to galaxy clustering. However, while the $k\nn$ distributions are statistically powerful characterizations the clustering of a set of galaxies (tracers), the final goal is to extract the cosmological parameters of interest in an unbiased manner. This can only be done after marginalizing over an appropriate set parameters that model the \textit{a priori} unknown connection between the underlying matter field, that is controlled by the cosmological parameters, and the galaxies that are observed. In other words, the tracer--matter connection needs to be understood in the language of these statistics in order to enable their application to data.

In this paper, we investigate the use of the HEFT framework as the model of the tracer--matter connection for $k\nn$ statistics. At the same time, this paper also represents the first application of the HEFT framework to any statistics beyond the $2$-point function. The paper is organized as follows: in Sec. \ref{sec:formalism}, we develop the formalism for predicting the $k\nn$-$\cdf$s of a set of discrete tracers of a clustered, continuous field in terms of the statistical properties of the field itself. We also provide a concrete implementation of the formalism for the case where the tracers are a random subset of particles from a cosmological $N$-body simulation, and the underlying field is the matter field at that redshift. In Sec. \ref{sec:halo}, we give a brief introduction to the HEFT framework, and outline how to compute $k\nn$ distributions of biased tracers from the ``weight'' fields generated by the HEFT formalism. In Sec. \ref{sec:results}, we choose a halo sample, whose mass range is roughly representative of the Luminous Red Galaxy (LRG) sample hosts at $z=0.5$, from simulations and fit its $k\nn$ distributions using the formalism from Sec. \ref{sec:halo}. We identify a set of scales over which HEFT is an accurate description of the tracer--matter connection for various $k\nn$ distributions. We also demonstrate that there are a set of bias values which can jointly model both the $2$-point clustering \textit{and} the $k\nn$ distributions of the halo sample. In Sec. \ref{sec:constraints}, we use a Fisher matrix formalism to estimate the potential gains in cosmological parameter constraints within the HEFT framework when $k\nn$ measurements are considered in addition to the traditional $2$-point measurements. We summarize the main points of our findings, and discuss certain interesting features of the study in Sec. \ref{sec:conclusions} .

\section{Formalism and calculation framework}
\label{sec:formalism}
In this section, we will lay out the formalism for making predictions for the $k\nn$-$\cdf$ measurements of discrete tracers from the properties of a continuous underlying field, assuming the tracers represent a Poisson sampling of the underlying field (see \cite{Banerjee_Abel} for details). Having laid out the formalism for a general field, we will demonstrate its application to the prediction of $k\nn$-$\cdf$ of a set of simulation particles with mean number density $\bar n$ given the continuous underlying matter field in Sec. \ref{sec:matter}.

The general setup is that of $N$ tracer (data) points, whose clustering we want to characterize, distributed over a total volume $V_{\rm tot}$. To compute the nearest neighbor distribution, the same volume is densely populated with a set of query points, and the distances to the $k$-nearest data point is measured from each query point. These distances, for a specific value of $k$, are then sorted to calculate the Empirical $\cdf$ of the distribution of distances to the $k$-th nearest neighbor data points. These measurements, while sensitive to higher order $N$-point functions, are computationally much cheaper than those typically used to measure information beyond $2$-point clustering. As shown in \cite{Banerjee_Abel}, the  value of the $\cdf$   of the $k$-th nearest neighbor distribution at radius $r$ is equal to the probability of finding more than $k-1$ tracers in a spherical volume $V=4/3\pi r^3$:
\eq{cdf_P_connection}{\cdf_{k}(r) = \mathcal P_{>k-1}(V) = 1 - \Big(\mathcal P_0(V) + ... + \mathcal P_{k}(V)\Big)\,,}
where $\mathcal P_k(V)$ is the probability of finding $k$ tracers in a volume $V$. Note that these probabilities are averaged over the entire volume of interest.

If the tracers represent a Poisson sampling of an underlying continuous field $\rho(\mbx)$, then the probability of finding $k$ tracers in a sphere of radius $r$ around a specific point $\mbx_i$ is given by
\eq{single_point}{\mathcal P_k(V)|_{\mbx_i} = \frac{\Big(M\left(\mbx_i,V\right)/m\Big)^k}{k!}\exp\Big[-M\left(\mbx_i,V\right)/m\Big]\,,}
where 
\eq{mass_definition}{M(\mbx_i,V) = \int _{|\mbx - \mbx_i|<R}\rho(\mbx) {\rm d}^3\mbx \,,}
or the total ``mass'' in the sphere of radius $R$ around point $\mbx_i$. $m$ represents the ``mass'' of each tracer. Note that the ratio of $M/m$ is dimensionless. The volume averaged probability $\mathcal P_k(V)$ can be expressed as 
\eq{average_prob}{\mathcal P_k(V) = \Big \langle \mathcal P_k(V)|_{\mbx_i}\Big \rangle \, ,}
where the average runs over all possible positions, $\mbx_i$ for the center of the sphere. As shown in \cite{Banerjee_Abel}, $\mathcal P_k(V)$ can also be formally expressed in terms of \textit{all} connected $N$-point functions, $\xi^{(N)}(\mbr_1,...,\mbr_N)$, of the underlying field,
\eq{npoint_expression}{\mathcal P_k(V) &= \frac{1}{k!}\Bigg[\Bigg(\frac{\rm d}{\rm d z}\Bigg)^k\exp\Bigg[\sum_{N=0}^{\infty}\frac{\bar n^N(z-1)^N}{N!}\nonumber \\ & \quad \quad \times \int_V...\int_V {\rm d}^3\mbr_1...{\rm d}^3\mbr_N\xi^{(N)}(\mbr_1...\mbr_N)\Bigg]\Bigg]_{z=0}\, ,}
where $\bar n$ is the mean number density of the tracers.

Combining Eqs. \ref{eq:cdf_P_connection}, \ref{eq:single_point}, and \ref{eq:average_prob}, the expression for value of the first three $k\nn$-$\cdf$ at some radius $r$ (or corresponding volume $V$) can be written as 
\eq{1nn_expression}{\cdf_{1\nn} & = 1 - \Big\langle \exp\left[-\lambda_i\right]\Big\rangle \\ \cdf_{2\nn} & = 1 - \Big\langle \exp\left[-\lambda_i\right]\Big\rangle - \Big\langle \lambda_i \exp\left[-\lambda_i\right]\Big\rangle \label{eq:2nn_expression}  \\ \cdf_{3\nn} & = 1 - \Big\langle \exp\left[-\lambda_i\right]\Big\rangle - \Big\langle \lambda_i \exp\left[-\lambda_i\right]\Big\rangle \nonumber \\ & \qquad- \frac 1 2  \Big\langle \lambda_i^2 \exp\left[-\lambda_i\right]\Big\rangle \label{eq:3nn_expression} \, ,}
where $\lambda_i = M(\mbx_i, V)/m$. From Eq. \ref{eq:npoint_expression}, it is clear that each $k\nn$-$\cdf$ measurement is sensitive to a different combination of the integrals of various $N$-point functions of the underlying field. 

\begin{figure*}
	\includegraphics[width=0.9\textwidth]{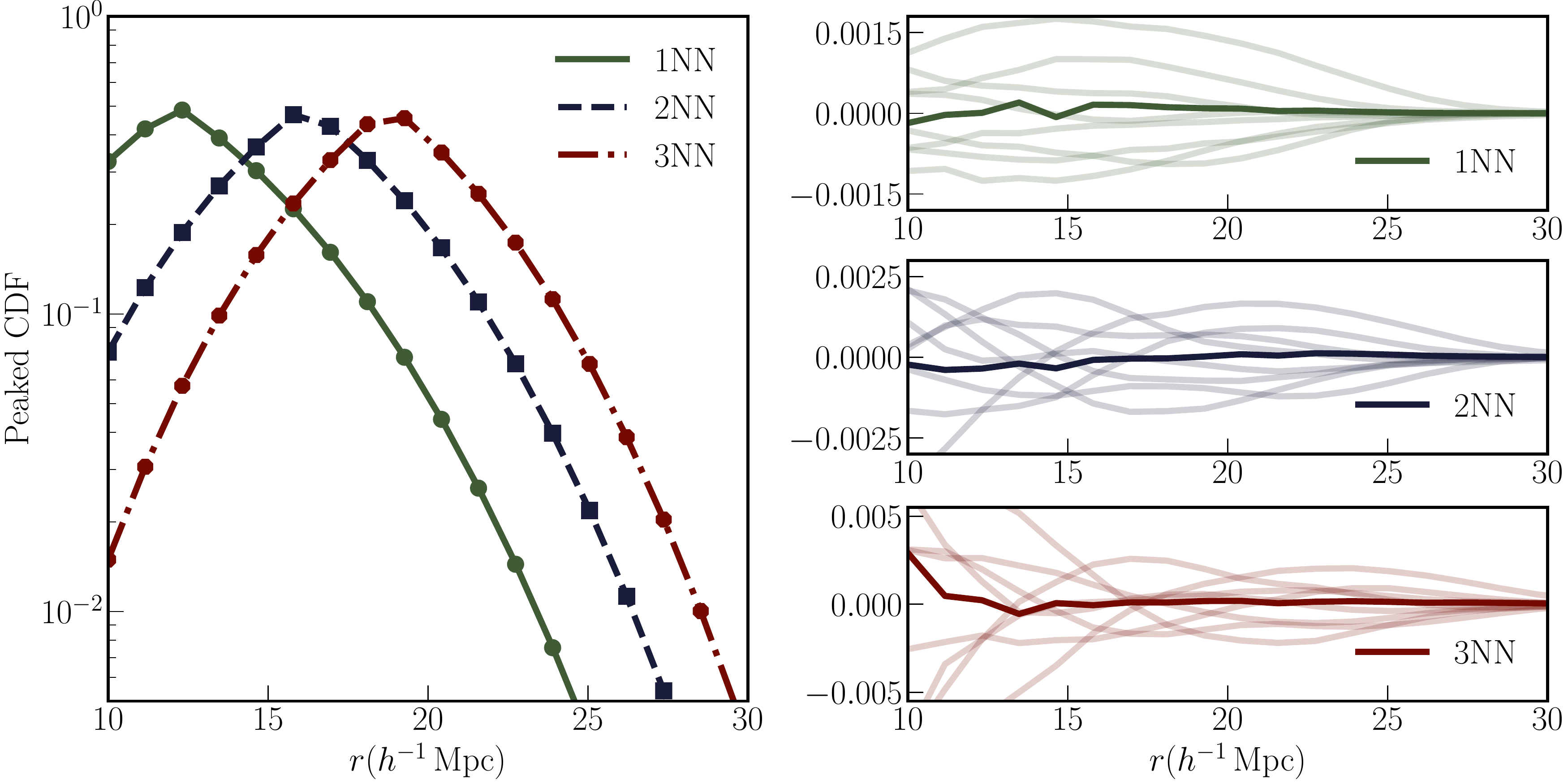}
	\caption{\textit{Left panel:} Peaked $k\nn$-$\cdf$s (defined in Eq. \ref{eq:peaked_cdf}) for $k\in \{1,2,3\}$. The different lines represent the measurements of these $k\nn$-$\cdf$s from a random set of $10^5$ simulation particles from an $N$-body simulation at $z=0.5$. The solid symbols represent the predictions of these $k\nn$-$\cdf$s computed from the smoothed matter field defined by depositing all simulation particles on a $1024^3$ grid, and following the procedure outlined in Sec. \ref{sec:matter}. The measurements and predictions are averaged over $8$ realizations at the same cosmology. The simulation volume is $(1\hgpc)^3$ and the total number of particles in each simulation is $512^3$.   \textit{Right panels}: Darker shaded lines represent the residuals between the measurements and the predictions in the left panel. The lighter shaded lines represent the variation in the measurements across realizations, i.e. sample variance. Each horizontal panel represents the residuals for a different $k$. Note that the $y$-axis stretch on each panels is different. Over the range of scales under consideration, the measurements and predictions are in very good agreement.}
	\label{fig:particles}
\end{figure*}

\citet{Banerjee_Abel_cross} demonstrated the extension of the $k\nn$ framework to the joint $\cdf$s of two sets of tracers - $N_1$ tracers of type $1$ and $N_2$ tracers of type $2$ distributed over a total volume $V_{\rm tot}$. The joint $\cdf_{k_1,k_2}$ is computed by measuring the distances to the $k_1$-th nearest neighbor data point from set $1$ and $k_2$-th nearest neighbor from set $2$ from a set of query points, and then considering the $\cdf$ of the larger of these two distances for every query point. The value of $\cdf_{k_1,k_2}$ at volume $V$ corresponds to the probability of finding more than $k_1$ data points from set $1$ \textit{and} $k_2$ data points from set $2$ in volume $V$. As shown in \citet{Banerjee_Abel_cross}, this joint $\cdf_{k_1,k_2}$ captures the auto-clustering of the two datasets, along with the cross-correlations between them. 

If the tracers represent Poisson samplings of two underlying fields defined by $\rho_1(\mbx)$ and $\rho_2(\mbx)$, then the probability of finding $k_1$ data points of set $1$ and $k_2$ data points of $k_2$ in volume $V$ around point $\mbx_i$ is given by
\eq{joint_Poisson}{\mathcal P_{k_1, k_2}(V)|_{\mbx_i} =& \frac{\Big(M_1\left(\mbx_i,V\right)/m_1\Big)^{k_1}}{k_1!}\exp\Big[-M_1\left(\mbx_i,V\right)/m_1\Big] \nonumber \\ & \times  \frac{\Big(M_2\left(\mbx_i,V\right)/m_2\Big)^{k_2}}{k_2!}\exp\Big[-M_2\left(\mbx_i,V\right)/m_2\Big]\, ,}
where $M_1$, $M_2$ are defined as in Eq. \ref{eq:mass_definition} for each field. The volume averaged $\mathcal P_{k_1, k_2}(V)$ is then defined as 
\eq{joint_volume_averaged}{\mathcal P_{k_1, k_2}(V) = \Big\langle \mathcal P_{k_1, k_2}(V)|_{\mbx_i} \big\rangle \, ,}
where once again, the average runs over all points $\mbx_i$ for the center of the sphere. In terms of the various connected $N$-point functions of the two fields, $\mathcal P(k_1,k_2)$ can be expressed as
\eq{joint_npoint_expression_1}{\mathcal P_{k_1,k_2}(V) = \frac{1}{k_1!}\frac{1}{k_2!}\Bigg[\Bigg(\frac{{\rm d}}{{\rm d}z_1}\Bigg)^{k_1}\Bigg(\frac{{\rm d}}{{\rm d}z_2}\Bigg)^{k_2} P(z_1,z_2|V) \Bigg]_{z_1,z_2=0}}
where 
\eq{joint_npoint_expression_2}{P(z_1,z_2|V) =& \exp\Bigg[\sum_{N_1=0}^\infty\sum_{N_2=0}^\infty \frac{\bar n_1^{N_1}(z_1-1)^{N_1}}{N_1!}\frac{\bar n_2^{N_2}(z_2-1)^{N_2}}{N_2!} \nonumber \\ &\times \int_V {\rm d}^3\mbr_1...{\rm d}^3\mbr_{N_1}{\rm d}^3\mbr^\prime_1...{\rm d}^3\mbr^\prime_{N_2} \xi^{(N_1,N_2)}\Bigg]\, ,}
where $\bar n_i$ represents the mean number density of each set of tracers. $\xi^{(N_1,N_2)}$ with $N_1\neq 0$, $N_2\neq 0$ represents all cross $N$-point functions formed from the two underlying fields.

Note that when the fluctuations in $\rho_1$ and $\rho_2$ are statistically independent, the above average breaks up into individual averages over the two fields. While the values of $\mathcal P_{k_1, k_2}(V)$ can be related to the joint $k\nn$-$\cdf$ for any general $k_1$ and $k_2$, in this paper, we will focus mainly on $\cdf_{1,1}$ which can be written as
\eq{CDF00}{\cdf_{1,1}(V) = 1 - P_0^1(V) - P_0^2(V)+P_{0,0}(V) \, }
where $P_0^1(V)$ and $P_0^2(V)$ represent the probability of finding $0$ particles in volume $V$ from set $1$ and $2$, respectively. $P_{0,0}(V)$ is given by setting $k_1=k_2=0$ in Eq. \ref{eq:joint_Poisson}. Therefore,
\eq{CDF00_explicit}{\cdf_{1,1}(V) =& 1 - \Big \langle \exp\left[ - \lambda_{1,i}\right]\Big \rangle  - \Big \langle \exp\left[ - \lambda_{2,i}\right]\Big \rangle \nonumber \\ & \quad + \Big \langle \exp\left[ - \lambda_{1,i} - \lambda_{2,i} \right]\Big \rangle \, ,}
where $\lambda_1 = M_1(\mbx_i,V)/m_1$ and $\lambda_2 = M_2(\mbx_i, V)/m_2$. The index $i$ runs over all possible sphere centers, with volume $V$, that can be defined in the total volume of interest.

\subsection{$k\nn$-$\cdf$s of the matter density field}
\label{sec:matter}

We will now provide a concrete application of the formalism presented above which will help demonstrate the calculation techniques we will subsequently use to predict the $k\nn$-$\cdf$s of biased tracers in Sec. \ref{sec:halo}. In this example, the continuous field that we consider is that of the matter field from a cosmological simulation at $z=0.5$, and we will use this field to predict the $k\nn$-$\cdf$s measured from a random subset of simulation particles. Since the late time matter field in an $N$-body simulation is computed from the particle positions themselves, the connection between the continuous field and the tracers via a local Poisson process is guaranteed. This is, therefore, an ideal example to demonstrate various features of the formalism.

To generate predictions and make measurements, we use $8$ of the fiducial simulations from the \textsc{Quijote} suite \footnote{https://github.com/franciscovillaescusa/Quijote-simulations} \citep{2020ApJS..250....2V}. These are $(1\hgpc)^3$ volumes with $512^3$ CDM particles, and we consider the $z=0.5$ snapshots, to be consistent with the choices in Sec. \ref{sec:halo}. For the ``tracers'', we select a random sample of $10^5$ particles, i.e. $\bar n = 10^{-4} (\hmpc)^{-3}$, from the full set of simulation particles, and measure the $k\nn$-$\cdf$s using this subset. The query points for the $k\nn$ measurements are placed on a $512^3$ grid. The matter field at the desired redshift is generated by depositing all the particles on to a $1024^3$ grid spanning the box. The default particle weight deposition scheme was the Cloud-in-Cell technique, but the results are robust to the use of other deposition schemes like Nearest Grid Point (NGP), or Triangular Shaped Cloud (TSC).

\begin{figure}
	\includegraphics[width=0.45\textwidth]{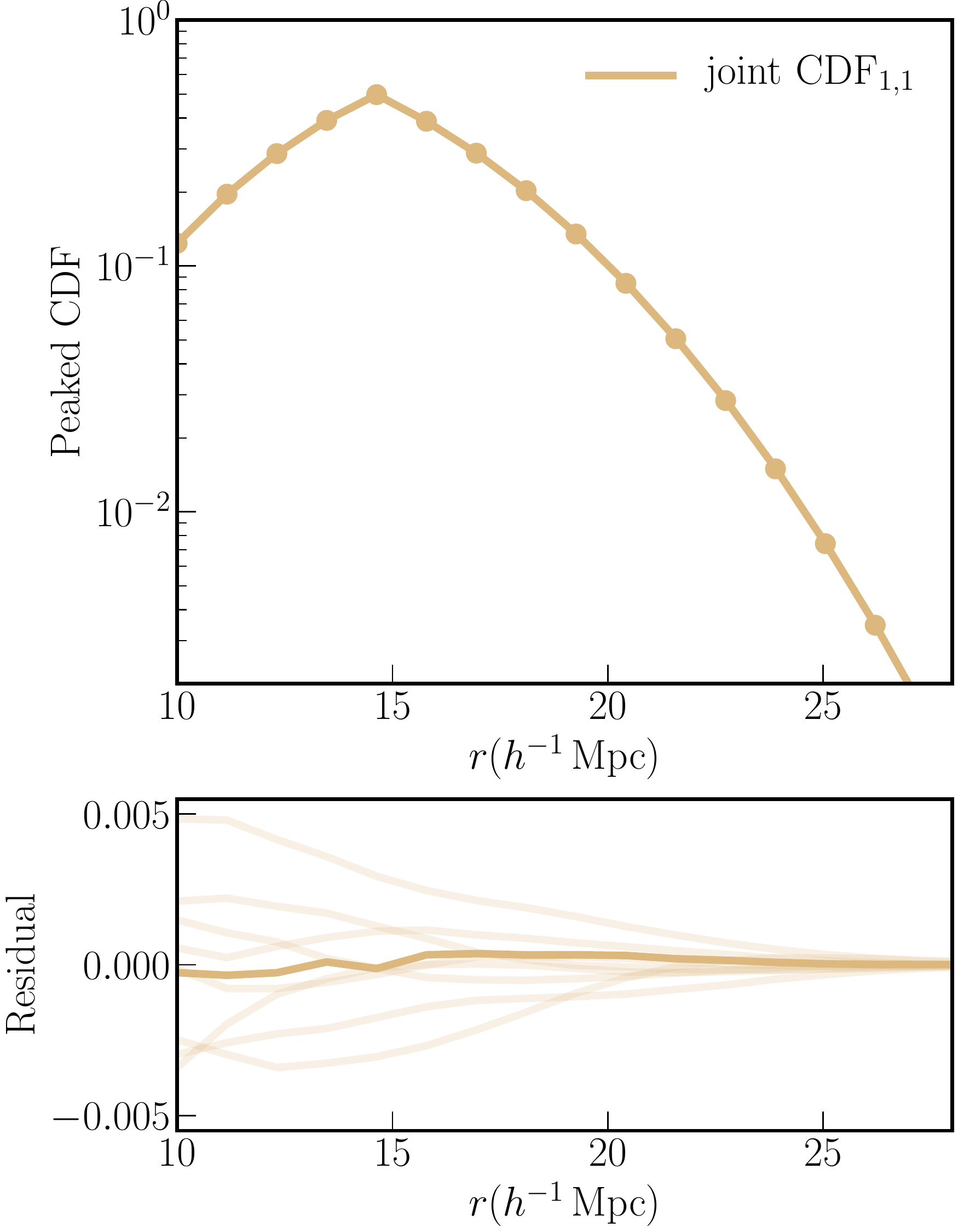}
	\caption{\textit{Top panel}: Peaked \textit{joint} $\cdf_{1,1}$ of two disjoint subsets of $10^5$ particles from a $512^3$ particle simulation over a $(1\hgpc)^3$ volume at $z=0.5$. The line represents the measurement from the particles. The solid data points represent the prediction for the $\cdf$ using the statistics of the smoothed matter field on a $1024^3$ grid, following the procedure in Sec. \ref{sec:matter}. Both the measurements and the predictions are averaged over $8$ realizations at the same cosmology. \textit{Bottom panel}: The darker shaded line represents the residual between the measurement and the prediction in the top panel. The lighter shaded line represents the variation in the measurement of the joint $\cdf_{1,1}$ from the $8$ different realizations compared to the mean. The joint $\cdf$ is sensitive to the cross-correlations between the two sets of tracers, and is well-predicted from the statistics of the underlying matter field over the full range of scales under consideration.}
	\label{fig:particle_joint}
\end{figure}

The set of scales we consider hereafter is set by two factors: the smallest scales are set by the range of applicability of the bias models we will consider in Sec. \ref{sec:halo}, i.e. $k_{\rm max} \sim 0.6\,  h{\rm Mpc}^{-1}$ or $r_{\rm min} \sim 10\,  \hmpc$. We also want to ensure that the scale of the grid on which the matter density is defined ($\sim 1 \hmpc$) is much smaller than the scales used in the calculation so that discreteness effects are minimized. The largest scale, $r_{\rm max}\sim 30 \hmpc$ is set by the number density of tracers -- the empirical $\cdf$ computed from the tracers is no longer a faithful representation of the true $\cdf$ once $\bar n V \gg 1$ (see \citet{Banerjee_Abel} for a discussion). Having fixed the scales, defined by $r_{\rm min}$ and $r_{\rm max}$, we now discuss the calculation procedure for obtaining the $k\nn$-$\cdf$s for $r_{\rm min}<r<r_{\rm max}$. We start by converting the density field $\rho$ on the $1024^3$ grid into an overdensity $\delta = (\rho - \bar \rho)/\bar \rho$, on the same grid. Next, the overdensity is smoothed on scale $r$. This is done in Fourier space using a top-hat filter. Each grid point is, therefore, assigned a smoothed overdensity $\delta_r$. Due to the choice of the filter, the mass enclosed in a volume $V = 4/3 \pi r^3$ around each grid point is simply $M_i  = \bar \rho V (1+\delta_{r,i})$, where $i$ now represents a grid index. Using this, we can rewrite Eqs. \ref{eq:1nn_expression}, \ref{eq:2nn_expression}, and \ref{eq:3nn_expression} as 
\eq{1nn_expression_matter}{\cdf_{1\nn} (r) &= 1 - \Big \langle \exp \left[-\bar \lambda\left(1+\delta_{r,i}\right ) \right] \Big\rangle \\ \cdf_{2\nn} (r)&= 1 - \Big \langle \exp \left[-\bar \lambda\left(1+\delta_{r,i}\right ) \right] \Big\rangle \nonumber \\ & \qquad -  \Big \langle \bar \lambda \left( 1+ \delta_{r,i}\right) \exp \left[-\bar \lambda\left(1+\delta_{r,i}\right ) \right] \Big\rangle \label{eq:2nn_expression_matter} \\ \cdf_{3\nn}(r) &= 1 - \Big \langle \exp \left[-\bar \lambda\left(1+\delta_{r,i}\right ) \right] \Big\rangle \nonumber \\ & \qquad -  \Big \langle \bar \lambda \left( 1+ \delta_{r,i}\right) \exp \left[-\bar \lambda\left(1+\delta_{r,i}\right ) \right] \Big\rangle \nonumber \\ & \qquad - \frac 1 2 \Big \langle \bar \lambda^2 \left( 1+ \delta_{r,i}\right)^2 \exp \left[-\bar \lambda\left(1+\delta_{r,i}\right ) \right] \Big\rangle\label{eq:3nn_expression_matter} \, ,}
where $\bar \lambda =  4/3 \pi r^3 \bar n$, and the average runs over all grid indices $i$. Since $\bar \lambda$ is a constant at fixed $r$, the above expressions require only the following averages to be computed on the grid: $\left \langle \exp\left[- \bar \lambda \delta_{r,i}\right]\right \rangle$, $\left \langle \delta_{r,i} \exp \left[- \bar \lambda \delta_{r,i}\right]\right\rangle$, and $\left \langle \delta_{r,i}^2 \exp \left[- \bar \lambda \delta_{r,i}\right]\right\rangle$. This is done by simply looping over the full grid.

The results of these calculations, averaged over the $8$ realizations, are presented in Fig. \ref{fig:particles}. The left panel shows the peaked $\cdf$s (PCDF) for the $k=1$, $k=2$, and $k=3$, where\begin{equation}
	{\rm PCDF(r) = \begin{cases}
		\cdf(r) \qquad &\cdf(r) \leq 0.5 \\
		1 - \cdf(r) \qquad &\cdf(r) >0.5 \, .
	\end{cases}}
	\label{eq:peaked_cdf}
\end{equation} 
We use the peaked $\cdf$ to plot the results since it clearly shows the trend on both tails, while the raw CDF always tends to $1$ at the right end, making it hard to discern small differences. The lines represent the $k\nn$ measurements from the $10^5$ tracer particles. The solid data points represent the prediction for the $k\nn$ measurements using Eqs. \ref{eq:1nn_expression_matter}, \ref{eq:2nn_expression_matter}, and \ref{eq:3nn_expression_matter}. We will use this somewhat counterintuitive convention throughout the paper, since the CDF measurements from data are made from $512^3$ measurements, one from each query point, producing a nearly continuous measurement of the empirical CDF. The predictions from the continuous field, on the other hand, are generated by smoothing the field over a finite set of scales. 
The darker shaded lines on each panel on the right represent the residual between the measurement and the prediction. The lighter shaded lines on each panel represent the variations in the measurements between the $8$ boxes. Note that the range of the $y$-axis is different in each panel. We find that over the scales of interest, the measurements and predictions are in good agreement, even though the raw values span almost two decades. The predictions and measurements  for $\cdf_{3\nn}$ below $15 \hmpc$ are slightly worse, but still well within the variation coming from different realizations. Note that higher $k\nn$ predictions are expected to be more susceptible to discreteness effects since they involve terms proportional to $\delta_{r,i}^k$ - any discreteness effect on $\delta_{r,i}$ is amplified when raised to power $k$, $k>1$.

Next, we turn to the calculations for predicting the joint $\cdf$s. For the $k\nn$-$ \cdf$ measurements, we simply choose two disjoint subsets of $10^5$ simulation particles from the set of all simulation particles. We then compute $\cdf_{1,1}$ using the process outlined in \citet{Banerjee_Abel_cross}. To make predictions for $\cdf_{1,1}$ from the underlying matter field, we use the fact that this situation corresponds to setting $\lambda_{1,i} = \lambda_{2,i} = \lambda_i$ in Eq. \ref{eq:CDF00_explicit}, i.e. the two underlying fields are completely correlated or identical. In terms of the smoothed matter overdensity field at scale $r$, Eq. \ref{eq:CDF00_explicit} can be rewritten as 
\eq{CDF00_explicit_matter}{\cdf_{1,1}(r) =& 1 - 2 \Big\langle \exp \left[- \bar \lambda \left(1+\delta_{r,i}\right)\right]\Big\rangle \nonumber \\ &\qquad + \Big\langle \exp\left[- 2\bar \lambda \left(1+\delta_{r,i}\right)\right]\Big\rangle \,.}
Therefore, in addition to the terms computed previously by averaging over the grid values, we also need to compute $\left\langle \exp \left[-2\bar \lambda \delta_{r,i}\right]\right\rangle$.

The results, once again averaged over $8$ realizations, are plotted in Fig. \ref{fig:particle_joint}. The upper panel plots the Peaked CDF - the line represents the measurement of the joint $\cdf_{1,1}$ from the particles, while the solid points represent the predictions from the continuous matter field. The bottom panel plots the residual between the prediction and the measurement using the darker shaded line. The lighter shaded lines represent the variation of the measurement itself over the $8$ boxes. Once again, we find that the measurements and predictions are in good agreement over the full range of scales, well within the variation from realization to realization.

Therefore, we have demonstrated that the $k\nn$-$\cdf$s for a set of Poisson tracers with number density $\bar n$ can be accurately predicted if the underlying continuous field is known, using the process outlined in this section. Furthermore, the choice of grid size, deposition scheme and other parameter choices are good enough to produce accurate predictions over the entire range of scales of interest. This serves as a useful reference while trying to determine the range of scales over which formalism presented in Sec. \ref{sec:halo} can be used to predict the $k\nn$-$\cdf$s of biased tracers.

\section{Hybrid EFT fields and nearest neighbor predictions for biased tracer fields}
\label{sec:halo}

In this Section, we give a brief introduction to Hybrid Effective Field Theory (HEFT) models of tracer clustering, and outline the calculational steps to predict the $k\nn$-$\cdf$s of a sample of tracers given the HEFT fields.

\subsection{Hybrid EFT and biased tracer fields}
\label{sec:hybrid_eft}
The Lagrangian bias models that are central to HEFT assume that at the initial, Lagrangian, coordinates $\bq$, the tracer density is related to the large-scale dark matter operators by a linear combination of all contributions allowed by Newtonian symmetries \citep{2016JCAP...12..007V}
\begin{align}
    \label{eqn:lagbias}
    \delta_X (\bq) &= F[\delta(\bq), s_{ij} (\bq)] \\
    &\approx 1 + b_1 \delta(\bq) + b_2 \left (\delta^2 (\bq) - \langle \delta^2 \rangle) \right ) + \nonumber  \\
    & \quad\quad b_{s^2} \left (s^2 (\bq) - \langle s^2 \rangle) \right ) + b_{\nabla^2} \nabla^2 \delta (\bq) + \cdots  \nonumber\\
    & \quad+ \epsilon (\bq). \nonumber
\end{align}
where we have expanded the functional $F$ to second order. Additionally, we have defined the \emph{traceless tidal tensor field} $s_{ij} (\bq) = \left ( \frac{\partial_i \partial_j}{\partial ^2} - \frac{1}{3}\delta_{ij} \right ) \delta (\bq)$. The field $\epsilon (\bq)$ describes the \emph{stochastic} contribution to the expansion. The stochastic term generally describes the fact the bias expansion describes a statistical relationship, and there can be deviations from this in any given realization. The stochastic fields are, by construction, uncorrelated with the bias operators $\langle \epsilon(\bq) \mathcal{O}_i \rangle = 0$. In the limit of $\epsilon(\bq)$ being distributed as Gaussian white noise, its auto-correlation is constant even after advection to late-times. \par
The Lagrangian coordinates $\bq$ are related to the current observed positions, $\bx$, through a coordinate transformation 
\begin{align}
    \bx (a) = \bq + \bPsi (\bq,a),
\end{align}
where $\bPsi$ is the \emph{displacement vector} that describes the motion of each Lagrangian fluid element from early times ($a \rightarrow 0$) to a late time with scale factor $a$\footnote{We will suppress the scale factor $a$ in subsequent equations so as to not over-load notation.}. In Lagrangian biasing theory \citep{Matsubara_2008} the late-time distribution of the tracer $\delta_X$ is then given by number conservation 
\begin{align}
    1 + \delta_X (\bx) = \int d^3 q F[\delta(\bq), s_{ij}(\bq)] \delta^D \left ( \bx - \bq - \bPsi(\bq) \right ) \, ,
	\label{eq:continuous_tracer_field}
\end{align}
where $\delta^D$ represents a Dirac delta function.
Assuming a prescription for $F[\delta, s_{ij}]$, a model for the displacements $\bPsi$ leads to a complete model of the summary statistics of the tracer distribution $\delta_X$. As mentioned, one can also use the $\bPsi$ from N-body simulations, as we will, leading to the classes of models called HEFT. 

In practice, HEFT corresponds to assigning the simulation particles different weights based on their Lagrangian, or initial, positions in the simulation volume. As the particles evolve under their collective gravity, they carry around this ``weight''. Note the distinction between the particle mass, which drives the gravitational evolution and the ``weight'' which is simply carried around by the motion of the particles - changing the weights do not alter the $N$-body dynamics. By adjusting the values of the bias parameters, and therefore, the weights associated with the particles,  the clustering of this tracer field can be adjusted to match the clustering of a target sample of halos or galaxies. It is important to note that the bias parameters $b_i$ considered in this paper are all scale-independent. \par
The tracer--tracer ($P_{XX}(k)$) and tracer--matter ($P_{Xm}(k)$) spectra, in HEFT, can then be written as sums of the cross-spectra of the advected fields in Eqn.~\ref{eqn:lagbias}. Defining the component spectra as $P_{ij} (k) \equiv \langle \mathcal{O}_i\mathcal{O}_j \rangle (k)$, for $i,j \in \{1, \delta(\bq), \cdots\}$, we then have
\begin{align}
    \label{eqn:p_hh} P_{XX} (k) &= \sum_{i,j} b_{i}b_{j} P_{ij} (k), \\
    \label{eqn:p_hm} P_{Xm} (k) &= P_{11}(k) + b_1 P_{1\delta}(k) + b_2 P_{1\delta^2}(k) \nonumber \\ & \qquad + b_{s^2} P_{1s^2}(k) + b_{\nabla^2} P_{1\nabla^2}(k),
\end{align}
where in the second line above we have explicitly written out the sum. The label `1' in any of the power spectra above represents the  the matter density field --- $P_{11}(k)$ is the power spectrum of the matter density field . The dependence of the bias parameters in HEFT on $N$-point functions is analytic, however this is not necessarily the case for other summary statistics such as $k\nn$-$\cdf$s. 

Crucially, HEFT models allow for the construction of field-level realizations (at low redshifts) of the different operators that contribute to the bias expansion, by simply summing over the relevant weights at the low redshift positions of the particles in the simulation. While most of the existing literature on HEFT has focused on the power spectrum as the summary statistic to quantify the clustering, the availability of the fields themselves mean that it is also possible to measure any other desired summary statistic, and assess the performance of the bias expansion beyond the traditional comparisons made at the level of the two-point correlation function. Given that $k\nn$-$\cdf$s measure various combinations of all connected $N$-point functions, their signals in HEFT are an ideal stress test of the Lagrangian bias formalism. \par

\begin{figure*}
	\includegraphics[width=0.75\textwidth]{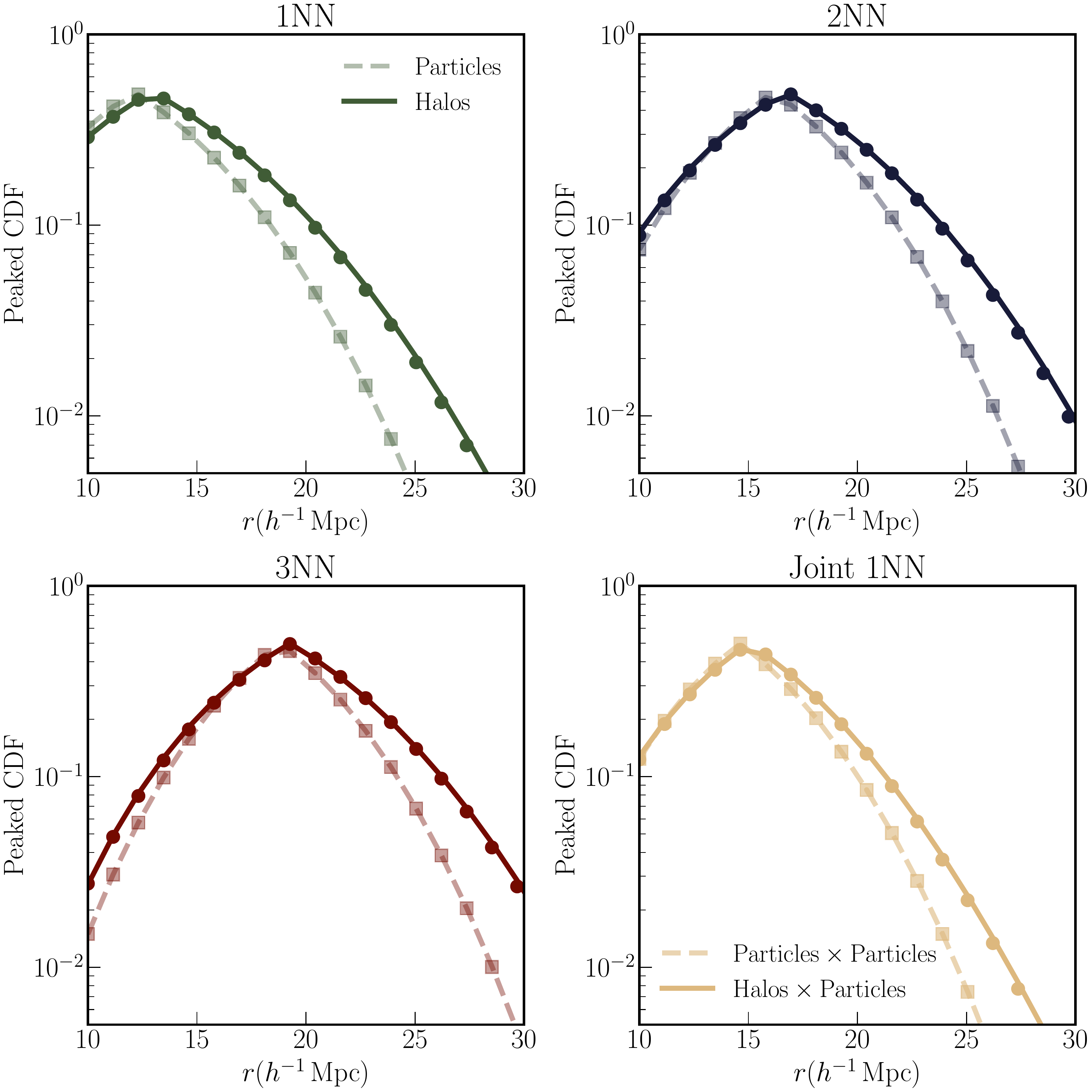}
	\caption{The solid lines in each panel represent the peaked CDF measurements at $z=0.5$ of various $k\nn$ distributions of the sample of $10^5$ halos, within the mass range  $\left(10^{13.26}-10^{13.5}M_\odot/h\right)$,  over a volume of $(1\hgpc)^3$ defined in Sec. \ref{sec:sim_halos}. The bottom right panel shows the measurement of the joint $\cdf_{1,1}$ between the halo sample and a random subset of $10^5$ simulation particles. The circular data points represent the prediction for the different $k\nn$-$\cdf$s from the smoothed tracer field (Sec. \ref{sec:sims}) defined by the best-fit values from the 2-point functions of this halo sample. For contrast, the dashed lines represent the $k\nn$ measurements for $10^5$ simulation particles instead of halos, while the solid squares represent the predictions for these $k\nn$-$\cdf$s using the smoothed matter field, instead of the smoothed tracer field. All measurements and predictions are averaged over $8$ realizations at the same cosmology. The statistics of the smoothed tracer field are able to capture the main features of the $k\nn$-$\cdf$s of the halos, even though the bias values were optimized to match the 2-point clustering only.}
	\label{fig:halos}
\end{figure*}

\subsection{Computing $k\nn$-$\cdf$s from the HEFT fields}
\label{sec:sims}

For a given set of bias parameters, $b_i$, the late-time continuous (overdensity) field $\delta_X(\mbx)$ is defined as in Eq. \ref{eq:continuous_tracer_field}. In practice, this field is defined on a grid --- in this case, a $1024^3$ grid --- by depositing the particle weights using a Cloud-in-Cell deposition scheme. To predict the values of the $k\nn$-$\cdf$s for tracers of this field at some radius $r$, we need the mean number density of the tracers $\bar n$, and the $\delta_X$ field smoothed on scale $r$, $\delta_{X,r}$. Given these ingredients, and the expressions in Eqs. \ref{eq:1nn_expression_matter}, \ref{eq:2nn_expression_matter}, and \ref{eq:3nn_expression_matter}, the $k\nn$-$\cdf$ values of the tracers can be evaluated by looping over the grid of overdensities. Any change in the values of the bias parameters $b_i$ lead to a change in $\delta_X$, which then propagates to a change in the $k\nn$ values when the relevant averages are preformed over the grid. To compute the predictions for the joint $\cdf_{1,1}$ between a set of tracers of the biased HEFT field, and a set of tracers of the matter field, we use the smoothed overdensities, smoothed at scale $r$ of both fields, $\delta_{X,r,i}$ for the HEFT field, and $\delta_{r,i}$ for the matter field:
\eq{joint_explicit}{\cdf_{1,1}(r) =& 1 - \left\langle \exp\left[- \bar \lambda_X\left(1+\delta_{X,r,i}\right)\right]\right\rangle  \nonumber \\ & \quad  - \left\langle \exp\left[- \bar \lambda_m\left(1+\delta_{r,i}\right)\right]\right\rangle \nonumber \\ & \quad + \left\langle \exp\left[- \bar \lambda_X\left(1+\delta_{X,r,i}\right) - \bar \lambda_m\left(1+\delta_{r,i}\right)\right]\right\rangle \, , } 
where $\bar \lambda_X = 4/3\pi r^3 \bar n_h$, $\bar \lambda_m = 4/3\pi r^3 \bar n$, and $\bar n_h$ and $\bar n$ are the mean number density of tracers of the HEFT field and the matter field respectively. As before, the averages represent averaging these expressions over every point on the grid, labeled by the index $i$.

We note here that in our calculations, we assume $\epsilon (\bq)$ is normally distributed with mean $0$. We also assume that it is not spatially correlated, i.e. the value of $\epsilon(\bq)$ at each location on the initial grid can be drawn independently from the values at other grid points. Note that, unlike the 2-point correlation functions which are sensitive only to the variance of $\epsilon(\bq)$, the various $k\nn$-$\cdf$s are formally sensitive to different moments of the distribution of $\epsilon(\bq)$. However, as shown in Sec. \ref{sec:Scales}, we find that for the halo samples considered in this paper, the effect of the $\epsilon(\bq)$ term on the fit is negligible. Therefore, in this paper, we stick to the assumptions of Gaussianity and lack of spatial correlations stated above, and leave the investigation of the effects of loosening these assumptions to a later study. The statement that the effect of $\epsilon(\bq)$ is negligible in this case is not to be conflated with the statement that we disregard shot-noise when treating the $k\nn$-$\cdf$s.


\section{Matching nearest neighbor clustering for a halo sample}
\label{sec:results}

In this section, we present the results of fitting the $k\nn$-$\cdf$s of a set of simulation halos using the HEFT formalism presented in Sec. \ref{sec:halo}. We compare the best-fit values of $b_i$ for the 2-point functions, and the $k\nn$-$\cdf$s, and show that over a set of scales, both sets of summary statistics can be fit with the same values of $b_i$.

\subsection{Defining the halo sample}
\label{sec:sim_halos}

For the analysis presented in this section, we once again use $8$ realizations from the fiducial cosmology of the \quijote simulation suite. We consider the snapshots at $z=0.5$ for our calculations, since this is a typical redshift that will be probed by various current and upcoming optical surveys. The halos are identified from the simulation particles using an FoF (Friends-of-Friends) halo finder. Since the number density $\bar n$ needs to be kept constant between the measurements and predictions, we choose $10^5$ halos from each simulation in the following way: we mass order the halos in the halo catalog, and throw out the most massive $10^5$ halos and consider the mass range defined by the next $10^5$ entries. This is done to remove the very massive halos whose clustering is not well captured by second order HEFT down to the scales of interest. We include \emph{all} halos in the box that fall within this mass range. Since the FoF masses are discrete (in units of the particle mass), this produces a catalog with slightly more than $10^5$ halos. Depending on the way the halo catalog is written out, especially when the halo list is ordered by coordinates, not performing this step could artificially distort the distribution of halos and this distortion is propagated into the catalog's summary statistics. Finally, we select $10^5$ halos at random from this catalog to ensure a fixed number density across boxes. The halos that make the final catalog fall in the mass range $\left(10^{13.26}-10^{13.5}M_\odot/h\right)$ whose clustering is well described by HEFT for $k\lesssim 0.6 h {\rm Mpc}^{-1}$ (see \cite{modichenwhite}). We will, therefore, concentrate on scales $k \lesssim 0.6 h {\rm Mpc}^{-1}$ or $r \gtrsim 10 \hmpc$ in the rest of the paper.

\begin{figure*}
	\includegraphics[width=0.8\textwidth]{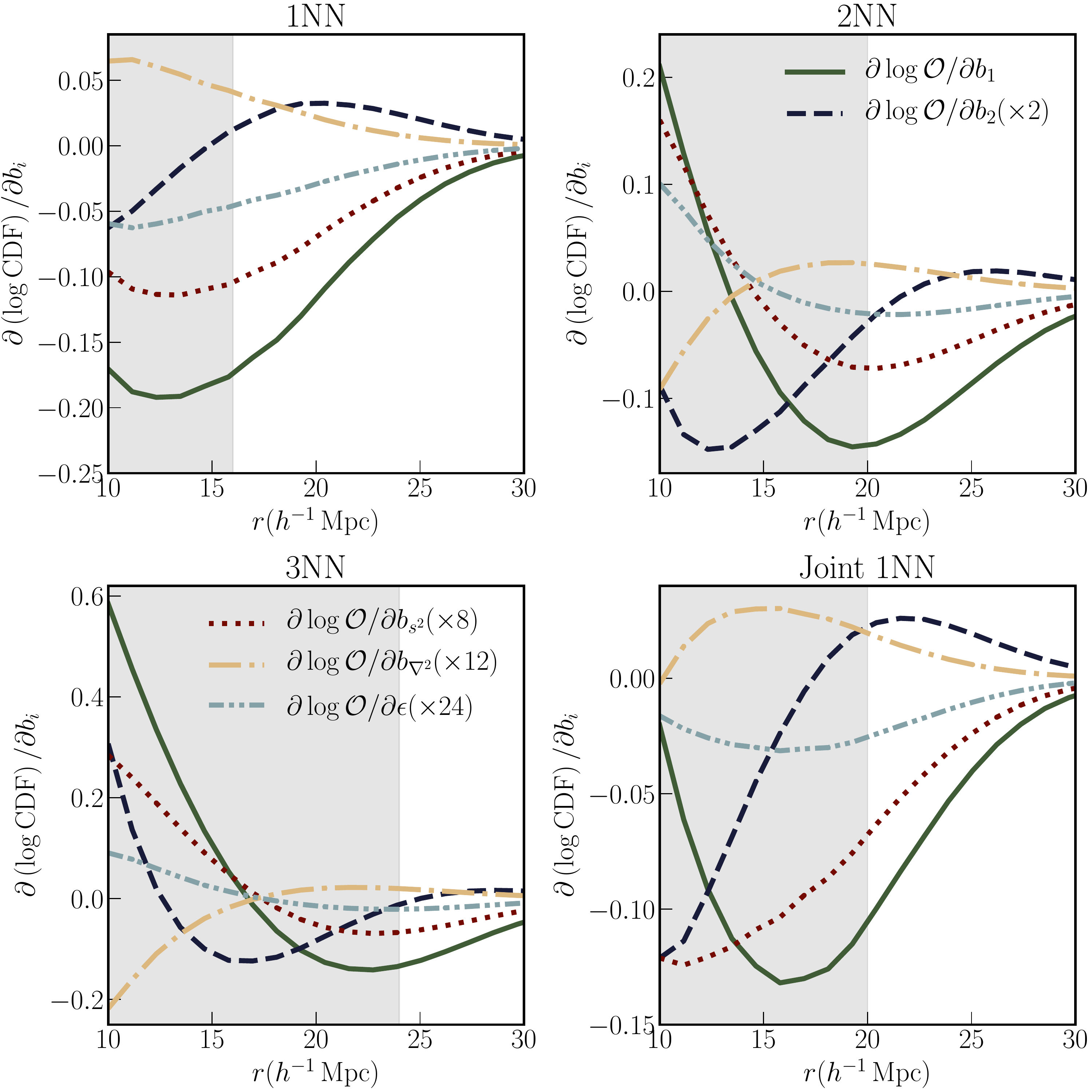}
	\caption{Logarithmic derivatives (around the best-fit values discussed in Sec. \ref{sec:2pt}) of different $k\nn$-$\cdf$s with respect to the second order Lagrangian bias parameters $b_1, b_2, b_{s^2}$ and $b_{\nabla^2}$. Each panel shows the response of a given $k\nn$ distribution, while lines of the same color represent the same bias parameter across the different panels. The grey shaded regions in each panel represent the scales where the HEFT predictions cannot accurately model the measurements, as discussed in Sec. \ref{sec:Scales}. Similar to the 2-point function, the predictions are most sensitive to changes in $b_1$. }
	\label{fig:deriv}
\end{figure*}

We compute the $k\nn$-$\cdf$s for this halo sample by placing query points on a $512^3$ grid, similar to the measurement done in Sec. \ref{sec:matter}. To compute the joint $k\nn$-$\cdf$ between the halo sample and the underlying matter field, which captures the cross-correlations of the two fields,we choose a random subset of $10^5$ particles from the full set of simulation particles, as representative tracers of the matter field. Our specific choice sets the number density of the halos and the matter tracers to be the same, but this is not necessary. In order to compute the relevant tracer fields at $z=0.5$, the linear fields (at $z_{\rm ini}=127$) relevant for the bias operators in Sec.~\ref{sec:hybrid_eft} are first computed using a modified version of the \textsc{2LPT}\footnote{https:
//cosmo.nyu.edu/roman/2LPT/} software. The bias operator fields are constructed and advected to $z=0.5$ using the \textsc{Anzu-fields}\footnote{https://github.com/kokron/anzu} package and the simulations's particle positions and noiseless linear density field as input. 


\subsection{Using best-fit bias values from 2-point clustering}
\label{sec:2pt}

We start by fitting the auto-correlation power spectrum $P_{hh}$, and the halo-matter cross-correlation spectrum $P_{hm}$ for the halo sampled defined above, using the procedure described in detail in \cite{kokron2021}. We use $k_{\rm max} = 0.6 h {\rm Mpc}^{-1}$ while performing the fit, to obtain the bias parameters $b_i$. In Appendix \ref{sec:kmax} we report the effect of how $b_i$ obtained from changing $k_{\rm max}$ affect the $k\nn$-$\cdf$s. We also note that, given the treatment of the shot noise term in \citet{kokron2021}, the variance of $\epsilon(\bq)$ is not determined separately from Poisson (or sampling) shot noise, and we set it to $0$ in this subsection. This yields a set of bias values $b_i$, with which we construct the $k\nn$-$\cdf$ predictions as outlined in Sec. \ref{sec:sims}. These predictions are compared to the actual measurements in Fig. \ref{fig:halos}. Each panel in the figure shows results for different $k\nn$ distributions - $k\in {1,2,3}$ for the halo sample, and the joint $\cdf_{1,1}$ between the halo sample and tracers of the matter density field. The darker solid line in each panel represents the $k\nn$ measurements. The solid circular points represent the predictions for each of the $k\nn$-$\cdf$s from the HEFT tracer field using the values $b_i$ obtained from the best-fit of the auto and cross power spectra. To emphasize the point that the late time tracer fields are generated from the same set of particles as the matter density field, but with different weights, we also plot the $k\nn$ measurements on particles (i.e. the same ones presented in Figs. \ref{fig:particles} and \ref{fig:particle_joint}) using the lighter dashed line in each panel, and the predictions from the matter using the square symbols. 

As can be seen clearly, the $k\nn$-$\cdf$ measurements of the particles and the halos are clearly different, evidenced by the difference between the solid and dashed lines in each panel. However, very importantly, the change from the square data points ---  computed from the matter density field deposited by the particles --- to the circular data points show that predictions for the $k\nn$ distributions of the halos computed from the HEFT tracer fields capture most of this change over the range of scales of interest. This is true even though the values of bias parameters were determined using the power spectra, while the $k\nn$-$\cdf$s are sensitive to integrals of all connected $N$-point functions of the tracer fields. Further, the bottom right panel shows that the same general agreement is also seen for the joint $\cdf_{1,1}$ between the halos and particles. This particular measurement is sensitive not only to all higher order $N$-point functions of the halo sample, but also to higher order $N$-point functions of the matter field, as well as all possible  $N$-point functions that can be defined in terms of various combinations of the two fields \citep{Banerjee_Abel_cross}. Therefore, the HEFT formalism can potentially fit not just the two-point clustering of the halo sample in terms of the two point functions of various advected Lagrangian fields, but also various higher order statistics. We explore this in more detail below.

\begin{figure*}
	\includegraphics[width=0.7\textwidth]{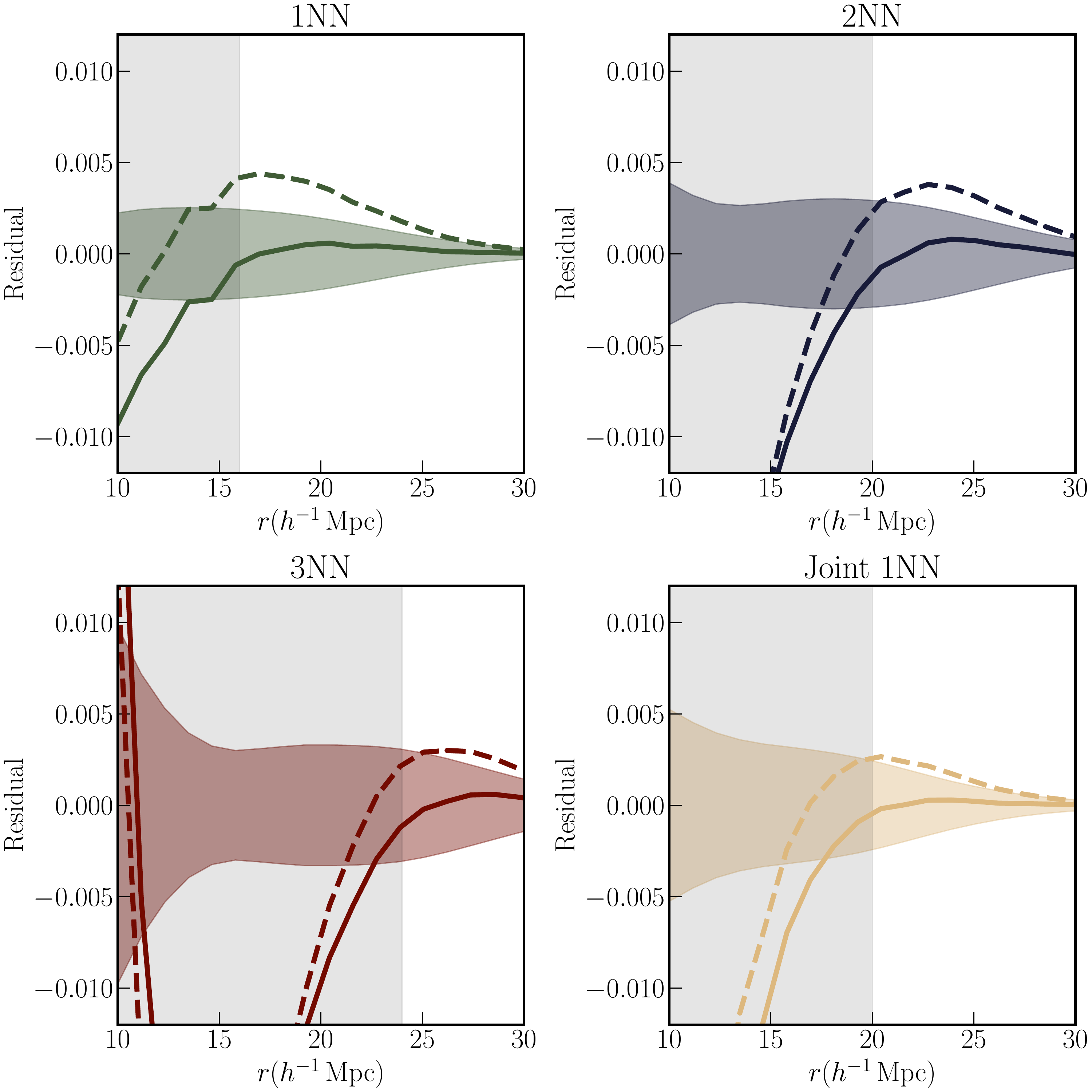}
	\caption{Residuals between the $k\nn$-$\cdf$ measurements of the halo sample from Sec. \ref{sec:sim_halos} and the predictions for the same generated using the smoothed HEFT fields. Each panel represents a different $k\nn$ distribution. The dashed lines represent the residuals in the case when the tracer fields were generated using the best-fit values from fitting the halo auto and cross power spectra. The solid lines represent the residuals when the bias values are optimized to fit the $k\nn$ distributions of the halos. All measurements and predictions are averaged over $8$ realizations. The shaded regions around $0$ represent the scatter in the halo $k\nn$ measurements from $100$ realizations. The grey shaded regions in each panel represent the scales where the HEFT predictions cannot accurately model the measurements. The solid lines show that there exist combinations of the bias parameters that produce a good fit to the halo $k\nn$-$\cdf$s over certain range of scales. The agreement extends down to the smallest scales for the $1\nn$ distribution $\sim 16 \hmpc$. See Fig. \ref{fig:residual_2pt} for the residuals with respect the power spectra measurements for the same set of bias parameters.}
	\label{fig:residual}
\end{figure*}

\subsection{Response to changes in bias  values}
\label{sec:Response}
\begin{figure*}
	\includegraphics[width=0.75\textwidth]{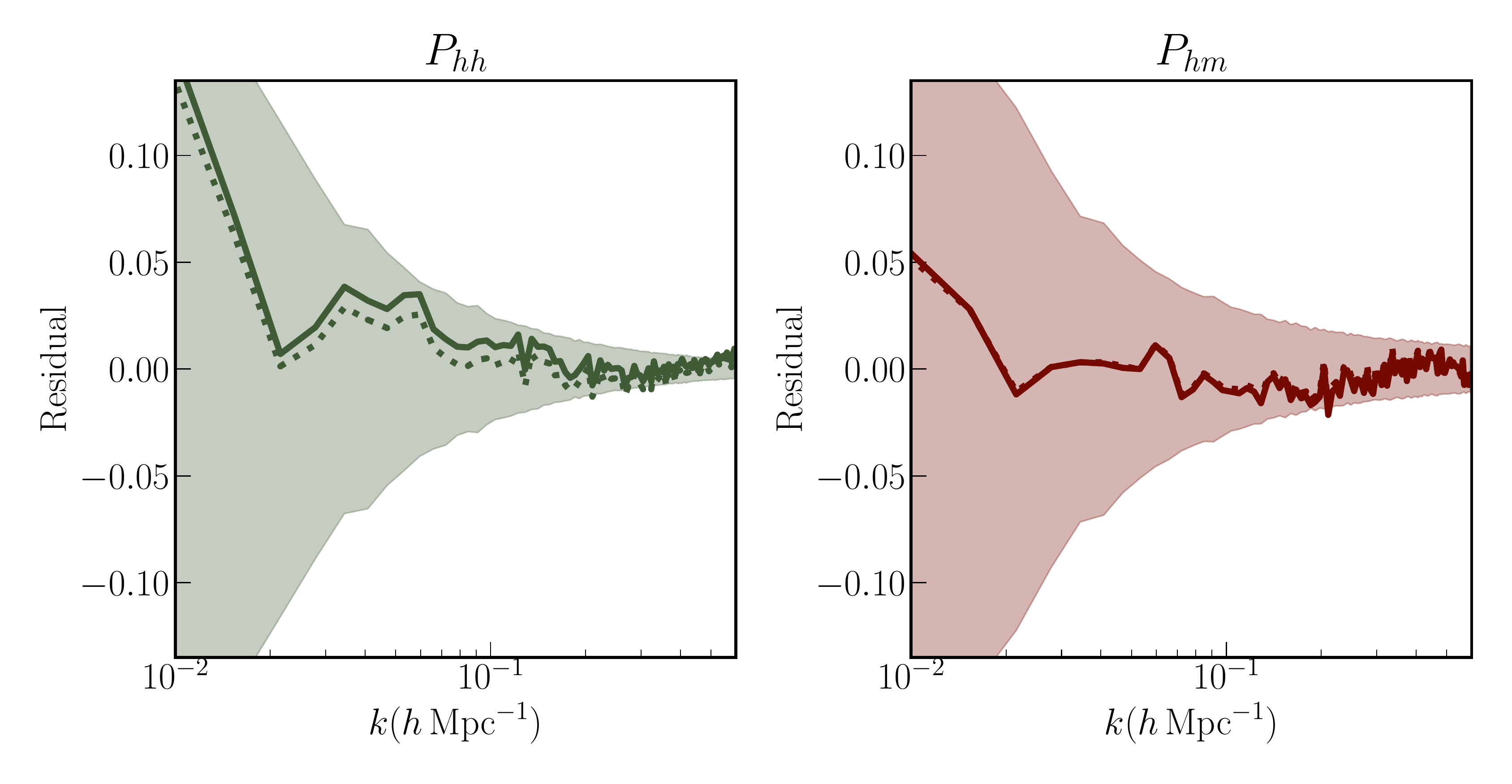}
	\caption{Residuals between the power spectra measurements of the halo sample from Sec. \ref{sec:sim_halos} and the predictions for the same generated using the HEFT fields. The left panel represents measurements and predictions of the halo auto spectrum, $P_{hh}(k)$. The right panel represents measurements and predictions of the halo-matter cross power spectrum $P_{hm}(k)$. The dotted lines represent the residuals in the case when the tracer fields were generated using the best-fit values from the halo auto and cross power spectra. The solid lines represent the residuals when the bias values are optimized to fit the $k\nn$ distributions of the halos, the same as the solid line in Fig. \ref{fig:residual}. All measurements and predictions are averaged over $8$ realizations. The shaded regions represent the scatter in the power spectra measurements from $100$ realizations. The difference between the two sets of curves is more pronounced in the left panel. Given the measurement uncertainties, the best-fit bias values obtained by fitting the $k\nn$ distributions are also provide a good fit to the power spectra measurements of the same halo sample.}
	\label{fig:residual_2pt}
\end{figure*}

We now investigate the response of the $k\nn$ distributions under consideration to changes in the values of the bias parameters around those obtained by fitting the two-point functions in Sec. \ref{sec:2pt}. To do this, we change the bias values, one at a time, by $5\%$ ($b_{i}^+$ and $b_{i}^-$) around the best-fit values, $b_i^{\rm fid}$. We recompute the predictions for the various $k\nn$-$\cdf$s with this new set of $b_i$, and express the change in terms of the logarithmic derivative of the quantities:
\eq{log_deriv}{\frac{\partial \log \mathcal O}{\partial b_i} \simeq \frac{1}{\mathcal O \left(b_i^{\rm fid}\right)}\frac{\mathcal O\left(b_i^+\right) - \mathcal O\left(b_i^-\right)}{b_i^+ - b_i^-} \, ,}
where $\mathcal O$ represents the various $k\nn$-$\cdf$s under consideration. 
Since we assumed $\langle \epsilon(\bq)^2\rangle = 0$ in the previous subsection, this parameter needs to be treated differently here. In this case, we use  one-sided finite difference for positive values of $\langle \epsilon(\bq)^2\rangle$ instead of a centered finite difference to approximate the derivative. To compute the one-sided finite difference, we evaluate the tracer fields and $k\nn$ predictions for $\sqrt {\langle \epsilon(\bq)^2\rangle} \in {0,0.1,0.2}$.

The results are presented in Fig. \ref{fig:deriv}. Each panel represents the change in the logarithmic derivative of a different $k\nn$-$\cdf$. The different lines on each panel represent the response of that particular $k\nn$-$\cdf$ prediction to a change in each of the bias parameters. Note that the logarithmic derivatives with respect to different bias parameters have been multiplied by different factors to make the changes visible on the same plot. As with the 2-point functions, the $k\nn$-$\cdf$ predictions are most sensitive to the linear bias value $b_1$, but all the bias parameters produce measurable changes on the different $k\nn$ distributions. Further, the same bias parameter, e.g. $b_1$, affects the different $k\nn$ distributions at different levels. This suggests that combining different $k\nn$ measurements in an analysis can help break degeneracies. We will explore this further in Sec. \ref{sec:constraints}.

\subsection{Best-fit to $k\nn$-$\cdf$s and range of scales of validity}
\label{sec:Scales}

\begin{figure*}
	\includegraphics[width=0.8\textwidth]{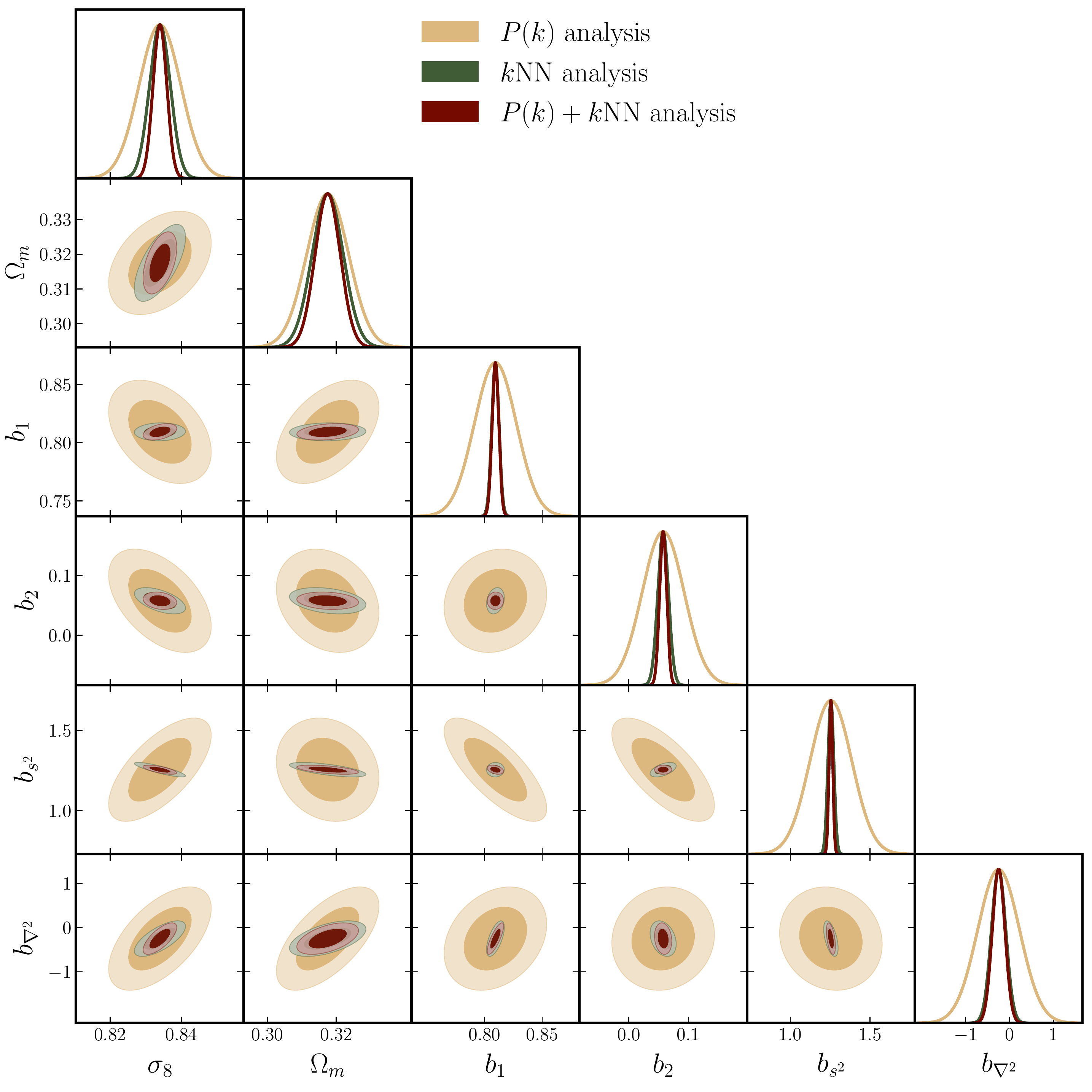}
	\caption{Results of the Fisher analysis presented in Sec. \ref{sec:constraints}. The yellow contours and standard Fisher errors represent the results when only the auto and cross power spectra of the HEFT fields are considered (up to $k_{\rm max} = 0.6 \ihmpc $). The green contours represent the standard Fisher errors from the analysis of the $k\nn$ measurements, over the range of scales discussed in the text. The marooon contours and standard Fisher errors represent the results of combining the $P(k)$ and $k\nn$ measurements. All forecasts assume the fiducial Quijote volume of $V=1 (\hgpc)^3$. For this sample, adding in the $k\nn$ measurements improves the constraints on $\sigma_8$ by a factor of $3$, and that on $\Omega_m$ by $\sim 60\%$ over the $P(k)$-only analysis. The HEFT bias parameters are also better constrained when $k\nn$ measurements are included.}
	\label{fig:contour}
\end{figure*}

Having established that changes in the bias parameters produce distinctive responses in the predictions for the $k\nn$-$\cdf$s for tracer of the HEFT fields, we now present the results of optimizing the bias parameters to fit the $k\nn$-$\cdf$s of the halo samples from the simulations. For the purposes of this first study, the bias parameters are changed by hand until the residuals discussed below are within the error bars. The results are presented in Fig. \ref{fig:residual}. We plot the residuals between the measurements of the halo $k\nn$-$\cdf$s and the predictions for the same from the smoothed tracers fields. The dashed lines represent the residuals for the case when the best-fit values from fitting the power spectra were used. The solid lines represent the residuals for the best-fit to the actual $k\nn$-$\cdf$ measurements. Each of the lines is produced by averaging over $8$ realizations. The colored bands around $0$ represent the variance of the $k\nn$-$\cdf$ measurements themselves from $100$ realizations at the same cosmology. We find that the best-fit bias parameters for the (auto and cross) power spectra of the halo sample are not optimal fits for the $k\nn$-$\cdf$s of the halo sample. A better fit can be produced by changing the values of the bias parameters, as seen by the solid lines in Fig. \ref{fig:residual}. Even in this case, however, depending on the $k\nn$ distribution under consideration, the prediction from the HEFT fields is a good fit, as defined by the spread of measurements across different realizations, to the halo measurements over a certain range of scales. The agreement persists down to the smallest scales for the $1\nn$ distribution (top left panel), $\sim 16 \hmpc$ . For the $2\nn$ distribution of the halos (top right panel), and the joint distribution of halos and tracers of the matter field (bottom right panel), the best-fit is a good match to the measurements down to $\sim 20 \hmpc$. For the $3\nn$ distribution (bottom left panel), the HEFT predictions are only a good fit at $\gtrsim 24 \hmpc$. These cutoff scales, below which the HEFT predictions do accurately model the $k\nn$ measurements are indicated by the shaded grey regions in each panel of Fig. \ref{fig:residual}. For context, the power spectra , $P_{hh}$ and $P_{hm}$, are well-fit by the HEFT prescription down to $k_{\rm max} \sim 0.6 \,h{\rm Mpc}^{-1}$ or $r \sim 10.5 \hmpc$. Since the higher $k\nn$ distributions are more sensitive to the higher density regions (see Eqs. \ref{eq:1nn_expression_matter}, \ref{eq:2nn_expression_matter}, and \ref{eq:3nn_expression_matter} for reference), these finding suggest that the HEFT fields do not fully replicate the clustering of halos in the highest density environments to this order in the Lagrangian bias expansion. On the other hand, for the $1\nn$-$\cdf$ is dominated by the distribution of the lowest density regions in the simulation, and the HEFT prescription does a good job of capturing the distribution down to scales closer to that for the power spectra. We note here that the $k\nn$ best-fits are consistent with $\sqrt {\langle \epsilon(\bq)^2\rangle} =0$.

In order for the HEFT model to model both the power spectra measurement, and the $k\nn$ measurements of a given tracer sample, it is crucial that the same set of bias values produce good fits for both sets of measurements. Therefore, we compute the predictions for the auto and cross power spectra given the bias values obtained from the $k\nn$ distribution fitting, and compare it to the measurements. In Fig. \ref{fig:residual_2pt}, we plot the residuals from this process using the solid lines. The dotted lines represent the residuals from the best-fit to the 2-point measurements themselves (as used in Sec. \ref{sec:2pt}). The left panel is for the auto power spectrum ($P_{hh}$) measurements , while the right panel is for the halo-matter cross spectrum ($P_{hm}$) measurements. The shaded region represents the spread in measurements from $100$ realizations. As can be seen, the solid lines are still a good fit to the measurements, given the error bars. Therefore, in the presence of these scale cuts, the same number of bias parameters, and equally importantly, the same values of these bias parameters in the HEFT framework can be used to model both the auto and cross power spectra of a set of tracers, as well as the $k\nn$ distributions of the same set of tracers. 

It is interesting to note that the change between the 2-point best fit and the $k\nn$ best fit -- i.e. the change from the dashed lines to the solid lines in Fig. \ref{fig:residual} -- is produced primarily by a change in the value of $b_{s^2}$. The value of $b_{s^2}$ changes by $\sim 13\%$ between the two. \citet{kokron2021} found that the tracer--tracer and tracer--matter spectra are least sensitive to varying $b_{s^2}$. It is therefore the hardest to constrain with only these statistics, as evidenced by the simulated likelihood analyses carried out therein. This can be understood, partially, by noting that the halo--matter spectrum is only sensitive to the parameter $b_{s^2}$ through the basis spectrum $P_{1s^2}(k)$ (Eqn.~\ref{eqn:p_hm}). This spectrum is orders of magnitude smaller than other basis spectra at all scales \citep[see fig.~1 of][]{kokron2021}. 

Given the results presented above the $k\nn$ distributions appear to be much more sensitive to $b_{s^2}$, and therefore should be able to constrain it much more strongly than $P(k)$ measurements can. This suggests using $k\nn$ measurements can be highly complementary to the traditional $P(k)$ or 2-point measurements, and combining them can help realize the full power of the HEFT framework in extracting cosmological parameter constraints from tracers of cosmological structure. We explore this aspect further in Sec. \ref{sec:constraints}.

\section{Fisher estimates for parameter constraints}
\label{sec:constraints}

\begin{table}
	\centering
	\caption  {$1$-$\sigma$ constraints on cosmological and HEFT parameters, from the $P(k)$-only, $k\nn$-only and $P(k) + k\nn$ Fisher analysis discussed in Sec. \ref{sec:constraints}.}
	\label{tab:constraints}
	\begin{tabular}{c|c|c|c}
		\hline
		\hline
		${\rm Parameter}$ & $\sigma_{P(k)}$ &$\sigma_{k\nn}$& $\sigma_{P(k)+k\nn}$\\
		\hline
		\hline
		$\sigma_8$ & 0.0058 & 0.0030 & 0.0019 \\
		$\Omega_m$ & 0.0061 & 0.0046& 0.0036 \\
		$b_1$ & 0.0182 & 0.0032 & 0.0029 \\
		$b_2$ & 0.0353 & 0.0091 & 0.0060 \\
		$b_{s^2}$ & 0.1314 & 0.0188 & 0.0116 \\
		$b_{\nabla^2}$ & 0.4787 &0.1677&  0.1460\\
		\hline		
	\end{tabular}
\end{table}

In this section, we demonstrate that using $k\nn$-$\cdf$ measurements in addition to the traditional 2-point function (or $P(k)$) measurements can improve constraints on both the HEFT model parameters, as well as the underlying cosmological parameters. To do this, we use the Fisher matrix formalism. The entries of the Fisher matrix ($\boldsymbol F$) are defined as 
\eq{Fisher}{\boldsymbol F_{\alpha\beta} = \sum_{i,j}\frac{\partial D_i}{\partial p_\alpha} \bigg[\boldsymbol C^{-1}\bigg]_{ij} \frac{\partial D_j}{\partial p_\beta}\, ,}
where $D_i$ represent the entries of the data vector under consideration (e.g. $P(k)$ values), $p_\alpha$ represent the parameters we are looking to constrain using the data vector. $\boldsymbol{C}$ represents the covariance matrix of the data vector at the fiducial set of parameters. We will consider two data vectors: the first is a data vector with the auto and cross power spectra of the HEFT fields with $k_{\rm max} = 0.6 \, h{\rm Mpc}^{-1}$. The second data vector is constructed by adding the $k\nn$-$\cdf$s of the HEFT fields to the first data vector. We note that we use the $k\nn$-$\cdf$s only on the scales where the HEFT fields provided a good fit to the halo sample in Sec. \ref{sec:Scales} - $16\hmpc$ for the $1\nn$-$\cdf$, and $20\hmpc$ for the $2\nn$-$\cdf$ and joint $\cdf_{1,1}$ between the HEFT field and the matter field. We discard the measurements on smaller scales, indicated by the grey shaded regions in Fig. \ref{fig:residual}. Since the range over which it is well modeled by HEFT is small, we do not use the $3\nn$-$\cdf$ in this analysis. We have shown in Sec. \ref{sec:Scales} that a single set of bias values can fit both the 2-point clustering and the $k\nn$ distributions of a tracer sample. Therefore, in this illustrative example, the HEFT fields and the clustering predictions are generated using the best-fit values used in Sec. \ref{sec:2pt} for the halo sample identified in Sec. \ref{sec:sim_halos}. In Appendix \ref{sec:redmagic}, we repeat the analysis for fiducial bias values that are representative of a different sample of tracers. The full set of parameters we use in this analysis are 2 cosmological parameters - $\sigma_8$ and $\Omega_m$, and the set of HEFT bias parameters - $b_1$, $b_2$, $b_{s^2}$ and $b_{\nabla^2}$. Note that in addition to these parameters, the $P(k)$ fit also uses a term controlling the  shot-noise amplitude. \par
To generate measurements over different realizations, we use the \textsc{Quijote} simulation suite - we estimate the covariance matrices for each data vector using $1000$ realizations of the fiducial cosmology in the \quijote suite. Specifically, to ensure that our covariance matrix properly captures the cross-correlations between the power spectra and the \knncdf s, for each box in the covariance suite we generate the HEFT field $\delta_{h}(\bx)$ from the fiducial bias values adopted in the forecast. The whole data vector is measured over this HEFT field. To compute the derivatives of the $\{P_{\rm hh}, P_{\rm hm}\}$, however, we use Eqns.~\ref{eqn:p_hh} and \ref{eqn:p_hm} and the measured basis spectra $P_{XY}$ for each box. Since a finite number of realizations are used to estimate the covariance matrix, we use the appropriate Hartlap factor \citep{2007A&A...464..399H} to correct for this. The cosmological derivatives, i.e. with respect to $\sigma_8$ and $\Omega_m$ are computed using $100$ realizations of the $\sigma_8 ^\pm $ and $\Omega_m ^\pm$ cosmologies from the same suite \citep{2020ApJS..250....2V}. The derivatives with respect to the bias parameters are computed using $100$ realizations of the fiducial cosmology, but by changing the bias values by $5\%$ around the fiducial values. The $1\sigma$ constraints on the parameters, labeled by $\alpha$, are obtained from the Fisher matrix as 
\eq{1sigma}{\sigma_\alpha = \sqrt{\big(\mathbf F^{-1}\big)_{\alpha\alpha}}\, .}
It should be noted that the standard Fisher error on parameter $\alpha$ represents the uncertainty after having marginalized over the other model parameters.

The results of the analysis are presented in Fig. \ref{fig:contour}. The $1\sigma$ constraints on individual parameters are summarized in Table \ref{tab:constraints}.The yellow contours and yellow standard Fisher errors in Fig. \ref{fig:contour} represent the results from the $P(k)$ Fisher analysis. The maroon contours and standard Fisher errors in Fig. \ref{fig:contour} represent the results from the combined $P(k)$ and $k\nn$-$\cdf$ analysis. We also plot the results from the $k\nn$-only analysis using the green contours and standard Fisher errors. We find that, for this halo sample, the constraints on $\sigma_8$ improve by roughly a  factor of $3$ when the $k\nn$-$\cdf$ measurements are added to the $P(k)$ measurements (maroon curves), while for $\Omega_m$, the improvement is somewhat smaller, $\sim 60\%$. This improvement in the cosmological parameter constraints can be summarized in terms of the Figure of Merit (FoM) in the $\Omega_m-\sigma_8$ plane, that is, the inverse of the area of the ellipse in Fig. \ref{fig:contour}. The FoM increases by a factor of $3.1$ when $k\nn$ measurements are added to the $P(k)$ measurements. By comparing the maroon and the green contours, it is clear that much of the gains are driven by the $k\nn$ distributions alone. For the bias parameters also, the standard Fisher errors are drastically reduced, for $b_1$ and $b_{s^2}$ in particular, when the $k\nn$ information is folded in. This is consistent with our finding in Sec. \ref{sec:Scales} where a change in $b_{s^2}$ was the primary difference between the $P(k)$ best-fit and the $k\nn$ best-fit. It is also worth noting that once again, the $k\nn$ measurements (green curves) drive most of the gain, even more so than for the cosmological parameters. Since the bias parameters are chosen to be scale-independent, the $k\nn$ measurements on relatively small scales are able to constrain the parameters precisely, and the $P(k)$ measurements do not add to the constraints. The cosmological parameters, on the other hand, produce unique changes on large scales that are not included in the $k\nn$ measurements, but are used in the $P(k)$ measurements. Therefore, combining the $P(k)$ and $k\nn$ measurements results in tighter constraints for the cosmological parameters than either set of measurements alone. The reduction in the uncertainties in the bias parameters, when $k\nn$ measurements are included, is exactly what drives the improvement in the cosmological constraints compared to the $P(k)$-only analysis. We explore this point further in Appendix \ref{sec:fixed_bias}, where we hold the bias parameters fixed, and compare the improvements on the cosmological parameters only. The cosmological and bias parameters can have somewhat degenerate effects on the 2-point clustering, but by combining with the $k\nn$ measurements, these degeneracies can be broken. We therefore conclude that there are potentially large gains to be made in terms of parameter constraints using the HEFT formalism when the $k\nn$ statistics are included to capture the clustering of tracers in addition to the 2-point clustering. 

While we have demonstrated this using a specific halo sample here, we show in Appendix \ref{sec:redmagic} that similar statistical gains from $k\nn$ measurements are expected to hold for another tracer sample relevant to various cosmological surveys - the \redmagic galaxy sample \citep{PhysRevD.98.042006}, provided that the modeling assumptions in this study extend to such a sample. Note that just the fact the constraints on all parameters improve when $k\nn$ measurements are considered in addition to the $P(k)$ measurements is not surprising in itself, since the addition of any statistics sensitive to the model parameters will improve the final constraints. However, it is worth pointing out again that the large improvements demonstrated above are possible because the $k\nn$ modeling does not require any additional model parameters compared to those needed to model the $P(k)$ measurements of the tracers. Indeed, this is one of the most important and useful conclusions from this study.

\section{Summary and Discussion}
\label{sec:conclusions}

In this paper, we have demonstrated how the Hybrid EFT (HEFT) formalism \citep{modichenwhite,kokron2021,2021arXiv210112187Z} can be used to model the $k$-nearest neighbor distributions \citep{Banerjee_Abel} of a set of biased tracers of the cosmological matter field over a range of scales in real space. To do so, we have developed the formalism to predict the $k\nn$-$\cdf$s of a set of Poisson tracers of a continuous field in terms of various averages of the underlying field smoothed on various scales. We have demonstrated the power of this formalism by computing the predictions for the $k\nn$-$\cdf$ of a random subset of simulation particles at $z=0.5$ in terms of matter density field at that redshift. We then use the same setup to match the $k\nn$-$\cdf$ measurements of a sample of halos from the \quijote simulations using the HEFT component fields and a set of values for the bias parameters. We show that even when the bias parameters are determined by fitting the auto and cross (with matter) power spectra of the halos, i.e., 2-point functions, the $k\nn$-$\cdf$ predictions from the resultant weighted HEFT fields provide a good approximation of the actual $k\nn$-$\cdf$ measurements from the halo sample. It is worth reiterating that the $k\nn$-$\cdf$ predictions depend on integrals of all connected $N$-point functions of the HEFT fields. We have explored how the $k\nn$-$\cdf$ predictions from the HEFT fields change in response to changes in the values of the bias parameters, and found that each bias parameter produces a unique response, especially when multiple nearest neighbor distributions are considered together. Crucially, we have then demonstrated that there exist a set of bias values, and by extension, a set of weighted HEFT fields, which can match both the 2-point clustering \textit{and} the $k\nn$ clustering of the halo sample, over a range of scales, with the $1\nn$ distribution most accurately predicted down to smaller scales. Finally, using a Fisher analysis, we show that including the $k\nn$ measurements, in addition to the 2-point measurements, can lead to much improved constraints on cosmological parameters within the HEFT framework. For $\sigma_8$ in particular, this improvement is roughly a factor of $3$. This paper, therefore, outlines a way of combining the HEFT modeling of biased tracers with the statistical power of the $k\nn$ formalism that can potentially extract much more information about the cosmological parameters of interest from quasi-linear scales probed by various surveys. We discuss some features of this study in further detail below.

While the $k\nn$ measurements are very sensitive to clustering information on scales below $\sim 30 \hmpc$ for the number density of tracers considered in this paper, these measurements do not fully capture the information on larger scales where the $\cdf$s all tend to $1$. Considering the $k\nn$ measurements by themselves would, for example, miss out on the information from the Baryon Acoustic Oscillation (BAO) scales ($\sim 100 \hmpc$). Therefore, combining the large-scale clustering information from the two-point function measurements with $k\nn$ measurements, which characterize the small-scale clustering accurately, represents an optimal data vector to measure clustering on all scales. Further, since the bias parameters affect the 2-point function and the $k\nn$ prediction in HEFT in different ways, the combination of these quantities also helps break various degeneracies.

The original formulation of the Hybrid EFT framework in \cite{modichenwhite} increased the range of scales over which the same set of Lagrangian bias parameters can be used to model the clustering of tracers compared to perturbation theory approaches, focusing on power spectra measurements to determine the range of scales. The model is able to match not only the auto power spectra of tracers, but also the tracer--matter cross spectrum. However, since the model explicitly uses the gravitational evolution of an $N$-body simulation particles, a characteristic feature of this model is that its primary output is a late-time nonlinear \textit{field} rather than any particular late-time summary statistic. Therefore, it is also possible to compute and compare any higher-order statistics between the HEFT predictions and a set of tracers. This is the first study which does so, by considering the $k\nn$ statistics of the tracer samples and the HEFT fields, which depends on integrals of \textit{all} connected $N$-point functions of the fields. The fact that the $k\nn$ statistics of the tracer sample can be modeled by the HEFT framework --- without expanding the number of free parameters needed --- over a range of scales, indicates that the HEFT formalism indeed works at the field level. This is further supported by the fact that the HEFT predictions also match the joint $\cdf$ between the tracers and matter. This particular $\cdf$ is sensitive to integrals of all possible $N$-point correlation functions between the tracer field and the underlying matter field, and can only be modeled correctly when both fields are evolved correctly. This suggests that while predicting $k\nn$ distributions, as was done in this paper, is particularly simple, the HEFT framework can also be applied to other higher order summary statistics that have been explored in the literature.

As shown in Sec. \ref{sec:Scales}, the range of scales over which the HEFT framework is able to fit the $k\nn$ distributions of tracers is not the same as the set of scales over which it provides a good fit to the power spectra, which are well-fit down to smaller scales. For the $1\nn$, the smallest scale that is accurately modeled, $r \geq 16 \hmpc$, is roughly comparable to that for the power spectrum. For higher $k\nn$ distributions, and for the joint distributions between the tracers and the matter field, the HEFT framework is only accurate at larger scales of $r \geq 20 \hmpc$. Given the statistical power of the $k\nn$ statistics, it will be interesting to explore if the modeling, especially for the higher and joint $\nn$ distributions, can be extended to smaller scales. One possibility is to include higher order bias terms --- currently terms up second order are included. Since the $N$-body dynamics themselves do not change with the inclusion of these terms, they simply represent more degrees of freedom in how the particle weights that make up the final HEFT fields can be distributed. However, it should be pointed that the number of terms allowed by symmetries keep increasing at higher orders, and thereby introducing many additional free parameters which need to be constrained by the data, in addition to the cosmological parameters of interest \citep{2018JCAP...09..008L,2018JCAP...07..029A}. Therefore, it needs to be investigated whether the gain in statistical power by modeling the $k\nn$ distribution from HEFT down to smaller scales is sufficient to offset the increase in the number of model parameters. For theoretical studies of higher order biasing in HEFT, though, the  statistical power of the nearest neighbor framework, along with the ease of calculation, should provide the right framework to isolate and understand the effects of these higher order terms on late-time clustering of biased tracers. 

The first step towards performing full cosmological analyses that harness the power of adding $k\nn$ measurements within the HEFT framework will be to predict these statistics as a joint function of the cosmological parameters and the bias parameters. Given that full $N$-body simulations are an essential component of the HEFT prediction, the emulator approach, which is increasingly being applied in the cosmological context, is best suited to this. \citet{kokron2021,2021arXiv210112187Z,boryana} have already the applied emulator approach to predicting the auto and cross power spectra of tracers as a function of cosmology and HEFT bias parameters. One of the natural extensions of this paper is to build a similar framework for $k\nn$ predictions, and will be explored in a future study.

Another notable extension of this work is to further investigate the impact of the stochasticity, $\epsilon(\bq)$, on our predictions. While we have found it sufficient to treat the field as Gaussian distributed with mean zero, some other recent studies \cite[see e.g.][]{2022MNRAS.tmp...77F} report the need for a non-Poisson stochastic term in the modeling of a related statistic - the PDF of projected galaxy counts. However, a direct comparison between results is difficult to make for multiple reasons. \citet{2022MNRAS.tmp...77F} concerns itself with a population of galaxies in projected space whereas we limit ourselves to a different halo sample in real space. Second, \citet{2022MNRAS.tmp...77F} uses a cylindrical collapse approach to the evolution of the underlying field, whereas this study makes use of $N$-body simulations for the same. Finally, the bias models themselves are somewhat different, since \citet{2022MNRAS.tmp...77F} restrict themselves to terms proportional to $b_1$ and $b_2$. Indeed, it is known that at the two-point function level \citep{2013PhRvD..88h3507B, 2017PhRvD..96h3528G} super-Poissonian stochasticity can arise from neglected bias operators. As we have pointed out in Fig.~\ref{fig:residual} the impact of slightly altering the tidal bias ($b_{s^2}$) in the residuals is significant, but this operator is neglected in \citet{2022MNRAS.tmp...77F}. 
Nevertheless, a more complete understanding and a thorough comparison of one-point PDF statistics, higher order Lagrangian bias models and stochasticity is of great interest, and we defer it for future work. In principle, if the resulting non-Poissonianity can be captured as a white noise field with a \emph{different} variance than expected, our treatment of stochasticity should still be sufficient. It is also possible to augment our current model by connecting the continuous field and the tracer (halo) sample through the two-parameter functional form used in \citet{2018PhRvD..98b3508F,2022MNRAS.tmp...77F}, rather than the single parameter Poisson sampling.

\section*{Acknowledgements}
The authors thank Andrey Kravtsov, Jessie Muir and Risa Wechsler for helpful comments on an earlier version of this manuscript. We thank the referee for insightful comments which helped improve the paper. This work was supported by the Fermi Research Alliance, LLC under Contract No. DE-AC02-07CH11359 with
the U.S. Department of Energy, the U.S. Department of Energy (DOE) Office of Science Distinguished Scientist Fellow Program, and the U.S. Department of Energy SLAC
Contract No. DE-AC02-76SF00515. Some of the computing for this project was performed on the Sherlock cluster. The authors would like to thank Stanford University and the Stanford Research Computing Center for providing computational resources and support that contributed to these research results. The \textsc{Pylians3}\footnote{https://github.com/franciscovillaescusa/Pylians3} and \textsc{nbodykit} \citep{Hand_2018} analysis libraries  were used extensively in this paper. We also acknowledge the use of the \textsc{GetDist}\footnote{https://getdist.readthedocs.io/en/latest/} \citep{2019arXiv191013970L} software for plotting.

\section*{Data Availability}

The simulation data used in this paper is publicly available at \url{https://github.com/franciscovillaescusa/Quijote-simulations}. Additional data is available on reasonable request.



\bibliographystyle{mnras}
\bibliography{ref} 




\appendix

\section{Changing scales used in the power spectra fit}
\label{sec:kmax}

\begin{figure*}
	\includegraphics[width=0.6\textwidth]{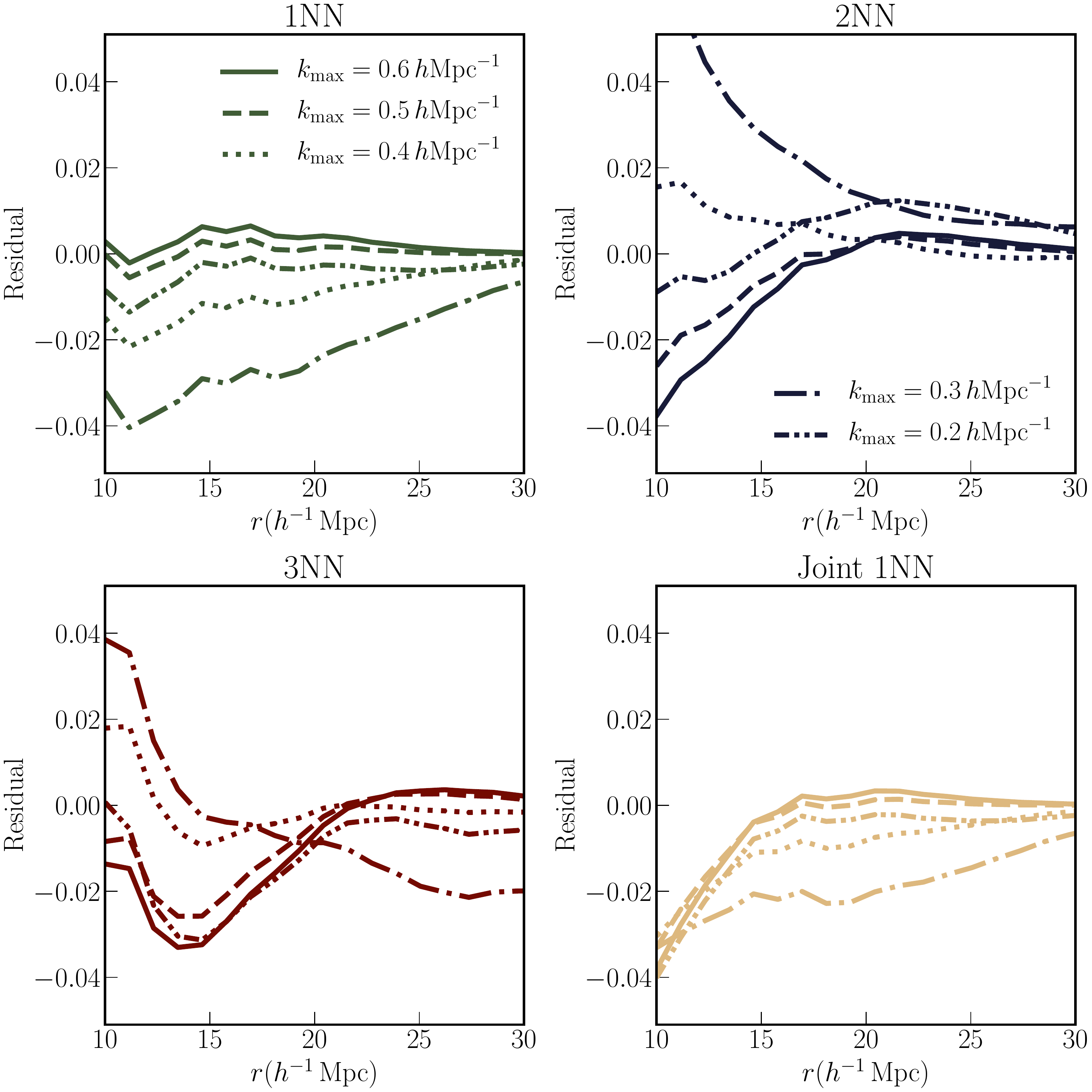}
	\caption{Plot of residuals between the measurements of various $k\nn$-$\cdf$s of the halo samples defined in Sec. \ref{sec:sim_halos} and the predictions for these quantities generated using the best-fit bias values from fitting the auto and cross power spectra, up to different $k_{\rm max}$. The solid line in each panel ($k_{\rm max} = 0.6\, h {\rm Mpc}^{-1}$) corresponds to the fiducial choice used in the rest of the paper. The other line styles correspond to lower choices for $k_{\rm max}$. For $k_{\rm max} \geq 0.5 \, h {\rm Mpc}^{-1}$, the predicted $k\nn$-$\cdf$s start to converge to the measurements over the range of scales discussed in Sec. \ref{sec:Scales}.}
	\label{fig:kmax}
\end{figure*}

For the results presented in Sec. \ref{sec:2pt}, the fitting to the auto power spectrum of the halos $P_{hh}$, and the halo-matter cross spectrum $P_{hm}$, was done up to $k_{\rm max} = 0.6\, h {\rm Mpc}^{-1}$. In this appendix, we illustrate the effect of different choices of $k_{\rm max}$ in the power spectra fitting on the predicted $k\nn$-$\cdf$ statistics via the best-fit bias values that each fitting procedure yields. We scan over the values of $k_{\rm max} \in \{0.2,0.3,0.4,0.5\} \, h {\rm Mpc}^{-1}$. We find the best-fit set of bias values for each $k_{\rm max}$, use this set to generate the HEFT fields, and use those fields as input to the calculation of the tracer $k\nn$-$\cdf$s. Note that irrespective of the $k_{\rm max}$ used in the fitting, we make predictions for the $k\nn$s over the same set of scales, $10\hmpc$ to $30\hmpc$. The results are presented in Fig. \ref{fig:kmax}. Each panel in the figure represents the residuals for a different $k\nn$-$\cdf$ between the measurements and the predictions. The solid lines in each panel correspond to the fiducial scenario of $k_{\rm max} = 0.6 \, h {\rm Mpc}^{-1}$. The other line styles correspond to other choices of $k_{\rm max}$. We find that for $k_{\rm max} \geq 0.5 \, h {\rm Mpc}^{-1}$, the predictions converge to those from the fiducial case, and are quite close to the actual measurements over the range of scales identified in Sec. \ref{sec:Scales}. For lower $k_{\rm max}$, the $k\nn$ predictions are not converged. They are also quite different (see e.g. the $k_{\rm max} = 0.4 \ihmpc$ curves) from the measurements, and this is true out to the largest scales on the plot. The lack of convergence is a direct result of the fact that the best-fit bias values from the 2-point functions themselves jump around as $k_{\rm max}$ is changed. In other words, for low $k_{\rm max}$, the total signal-to-noise is too low to effectively constrain all the HEFT model parameters. It is only once higher $k_{\rm max}$, and therefore higher signal-to-noise, values are used in the fit that the two-point measurements by themselves constrain the model parameters to a high degree of precision. The fact that the result converges in terms of the $k\nn$ measurements also suggests that when higher $k$ values, i.e. above $k=0.5 \, h {\rm Mpc}^{-1}$, are used in the fit, even though the fit is to 2-point functions only, the HEFT model and the bias values correctly capture a large fraction of the evolution of the underlying field itself, since the $k\nn$ predictions depend on various combinations of integrals of all $N$-point correlation function of the HEFT fields.

\section{Fisher constraints at fixed bias values}
\label{sec:fixed_bias}

\begin{figure}
	\includegraphics[width=0.45\textwidth]{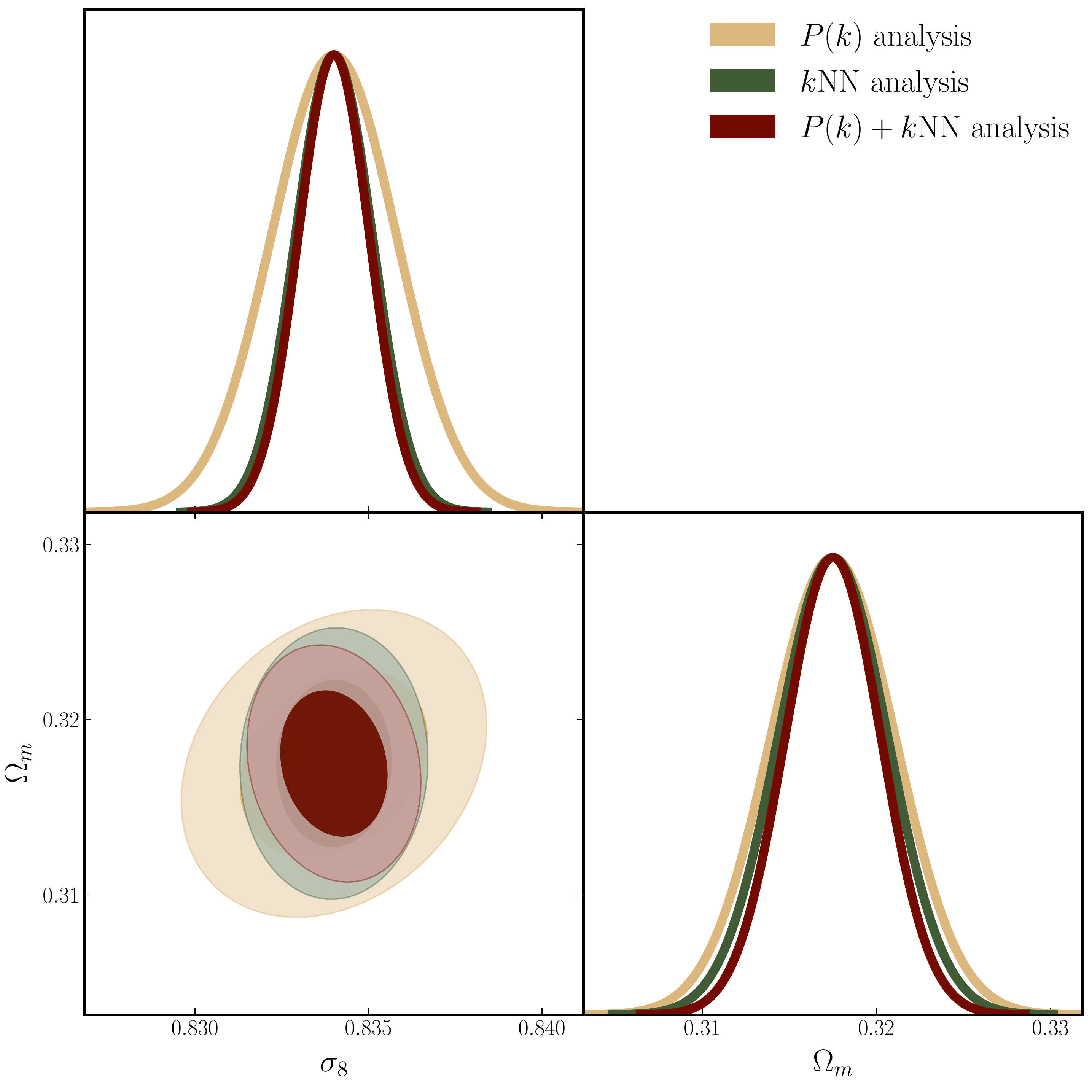}
	\caption{Fisher analysis results when the bias parameters are held fixed. The yellow curves represent the results from a $P(k)$-only analysis, the green curves represent the results of a $k\nn$-only analysis, while the maroon curves represent the results from a combined analysis. The constraint on $\sigma_8$ improves by $\sim 80\%$, while those on $\Omega_m$ improve by $\sim 20\%$ in the joint analysis compared to the $P(k)$-only analysis.}
	\label{fig:fixed_bias}
\end{figure}

In this appendix, we study the improvement in cosmological parameter constraints when we fix the values of the bias parameters in our model. This helps isolate the part of the improvement in cosmological parameter constraints driven by the greater sensitivity of the $k\nn$ measurements to the bias parameters. We repeat the calculations presented in Sec. \ref{sec:constraints}, varying the cosmological parameters around the fiducial values, but not the $b_i$. The results are presented in Fig. \ref{fig:fixed_bias}. We find that for $\sigma_8$, the constraints improve by $\sim 80\%$ in the joint analysis (maroon curves) compared to the $P(k)$-only analysis (yellow curves). On the other hand the constraints on $\Omega_m$ improve by only $\sim 20\%$. Note that these improvements, from the addition of the $k\nn$ measurements to the analysis, are smaller than those obtained in Sec. \ref{sec:constraints}, where the $\sigma_8$ constraint improved by a factor of $\sim 3$, and the $\Omega_m$ constraint improved by $\sim 60\%$. This suggests that the $k\nn$ measurements are crucial in breaking the degeneracies between the bias parameters and the cosmological parameters in the full analysis. It should be noted that we compare the relative improvements, and not the absolute values of the constraints, since the absolute constraints (for any of the data vectors) are tighter in the analysis with bias parameters held fixed, as opposed to when they are marginalized over, as is done in Sec. \ref{sec:constraints}.

\section{Fisher constraints for a different tracer sample}
\label{sec:redmagic}

\begin{figure*}
	\includegraphics[width=0.7\textwidth]{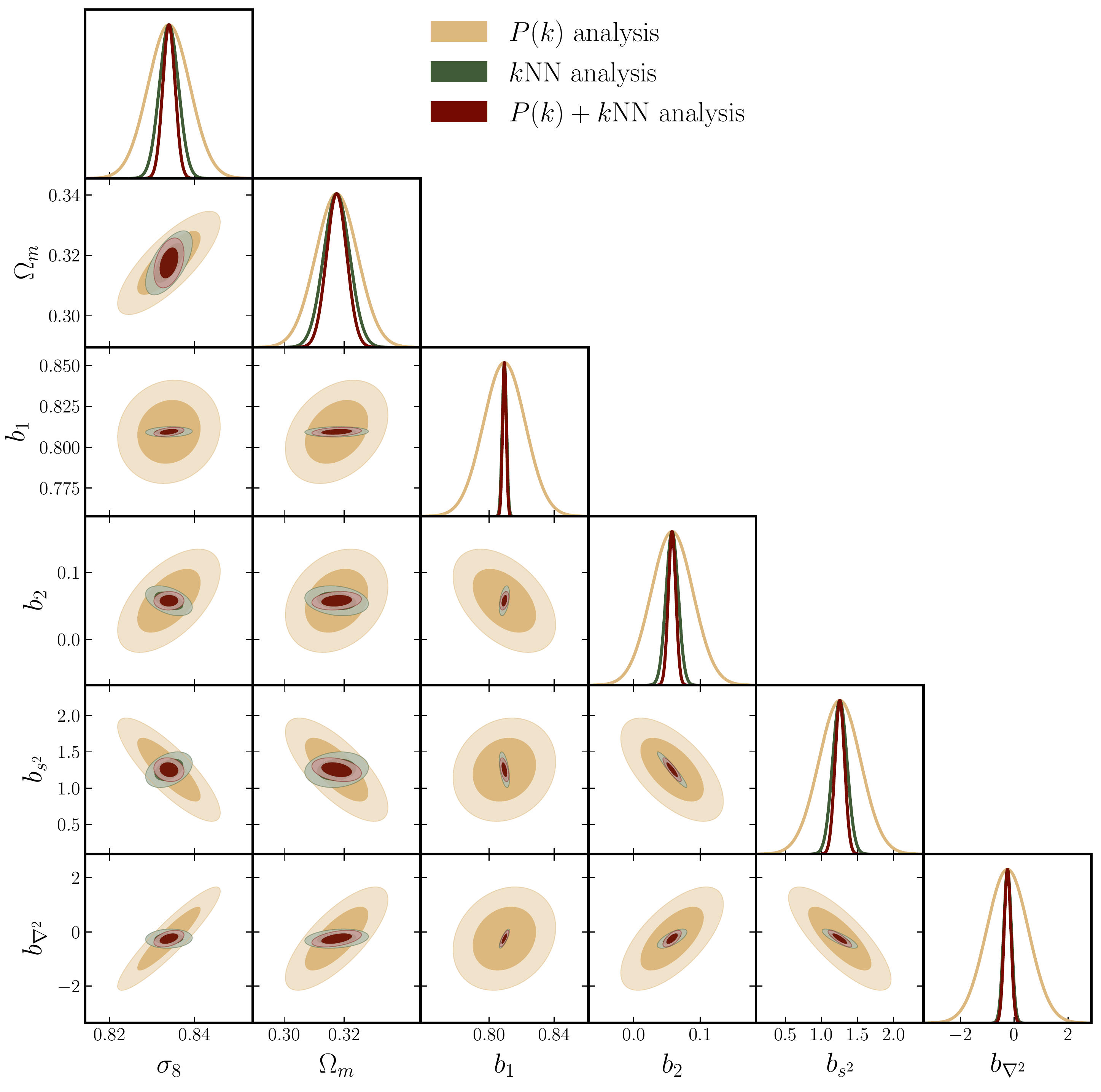}
	\caption{Results of the Fisher analysis of a \redmagic-like sample, as discussed in Appendix \ref{sec:redmagic}. The yellow contours and standard Fisher errors represent the results for the $P(k)$-only analysis. The green contours and standard Fisher errors represent the results from the analysis of the $k\nn$ distributions only, while the maroon contours and Fisher errors represent the results when the two data vectors are combined. For this sample, the constraints on the cosmological parameters, $\sigma_8$ and $\Omega_m$ improve by more than a factor of $2$ when $k\nn$ measurements are added to the $P(k)$ measurements. The bias parameters are also more tightly constrained in this case.}
	\label{fig:redmagic}
\end{figure*}

\begin{table}
	\centering
	\caption  {$1$-$\sigma$ constraints on cosmological and HEFT parameters, from the $P(k)$-only, $k\nn$-only, and $P(k) + k\nn$ Fisher analysis discussed for the \redmagic sample discussed in Appendix \ref{sec:redmagic}.}
	\label{tab:redmagic}
	\begin{tabular}{c|c|c|c}
		\hline
		\hline
		${\rm Parameter}$ & $\sigma_{P(k)}$ & $\sigma_{k\nn}$&$\sigma_{P(k)+k\nn}$\\
		\hline
		\hline
		$\sigma_8$ & 0.0049 &0.0022& 0.0014 \\
		$\Omega_m$ & 0.0070 &0.0044& 0.0034 \\
		$b_1$ & 0.0129 &0.0014& 0.0013 \\
		$b_2$ & 0.0315 &0.0093& 0.0057 \\
		$b_{s^2}$ & 0.2907 & 0.102 & 0.0669 \\
		$b_{\nabla^2}$ & 0.7799 & 0.148 & 0.1262 \\
		\hline		
	\end{tabular}
\end{table}

In Sec. \ref{sec:constraints}, we presented the improvement in constraints on cosmological and bias parameters for a set of halos (defined in Sec. \ref{sec:sim_halos}) which could be directly identified in the simulations that were used (the fiducial resolution of the \quijote suite). However, since the mass cuts and number density of the sample were arbitrary choices, we expect the HEFT model to also predict the 2-point and $k\nn$ clustering for other tracer samples on roughly the same set of scales. In this Appendix, we explore the possible improvements in parameter constraints, when the $k\nn$ measurements are considered in addition to the 2-point measurements, for a set of tracers that are relevant for many cosmological surveys. The tracers we now consider are the \redmagic sample of Luminous Red Galaxies use in both clustering and lensing studies \citep{PhysRevD.98.042006}. The Halo Occupation Distribution (HOD, see \citet{2005ApJ...633..791Z}) of this sample has been characterized in \cite{2016MNRAS.461.1431R,2017MNRAS.465.4204C}. \citet{kokron2021} demonstrated that the HEFT formalism is able to fit the 2-point clustering of such a sample, using the high resolution \textsc{Unit} simulations\footnote{http://www.unitsims.org/} \citep{2019MNRAS.487...48C} to populate halos with the HOD parameters from \cite{2017MNRAS.465.4204C}. These parameters are also consistent with more recent studies of the \redmagic HOD \citep{2021arXiv210608438Z}. Since the central occupation is significant for halos below the mass resolution of the \quijote suite, these simulations cannot be directly populated with the \redmagic  HOD, and therefore a rigorous study for such a sample is outside the scope of this paper, and will be done in future work. However, we can take the best-fit bias values of the HEFT model from \cite{kokron2021} and repeat the Fisher matrix calculation presented in Sec. \ref{sec:constraints}. We note that while the \redmagic sample is selected due to its reliable photometric redshift properties, the tests carried out below are performed purely in real space and without taking photometric redshift errors into account. That is, we are interested in the more fundamental question of potential improvements in constraints to be made, under the assumption that HEFT is a suitable model to describe the statistics of such galaxies. This is partially due to the \redmagic galaxies existing at separate fiducial points in the space of bias parameters. A full joint $P(k)$ + \knncdf test of \redmagic galaxies placed in a realistic simulation such as the \textsc{Buzzard} \citep{derose2019buzzard} suite is an interesting study but beyond the scope of this paper.

We retain the assumptions made in our $k\nn$ modeling from Sec. \ref{sec:halo}, such as Poisson sampling, even though the form of the HOD can, in principle, break this assumption on small scales. We use the same range of scales as in Sec. \ref{sec:Scales}, and for the sake of simplicity, use the same number density, $10^{-4}(\hmpc)^{-3}$, as in the rest of the paper. The native number density of the \redmagic galaxies is higher than this, so the sample considered here can be thought of as a random downsampling of a \redmagic-like sample. The scales that are included for each of the summary statistics are the same as those discussed in Sec. \ref{sec:constraints}.

The results are presented in Fig. \ref{fig:redmagic} and the $1$-$\sigma$ errors on the parameters are summarized in Table \ref{tab:redmagic}. The yellow contours and curves correspond to the data vector where only the auto (galaxy-galaxy) and cross (galaxy-matter) power spectra of the HEFT fields are considered. The maroon contours and curves correspond to the data vector where the $k\nn$ measurements are added. We also plot the results of the $k\nn$ analysis using the green contours and standard Fisher errors. In this case, the improvements on both cosmological parameters, $\sigma_8$ and $\Omega_m$ are pronounced - they both improve by more than a factor of $2$ when the $k\nn$ measurements are included. In terms of the FoM in the $\Omega_m-\sigma_8$ plane, the improvement is by a factor of $3.46$. The standard Fisher errors on the bias parameters also improve with the inclusion of $k\nn$ measurements, as was the case in the halo sample in Section~\ref{sec:halo}. Therefore, if the HEFT modeling of $k\nn$ distributions extend to such a sample, used by most cosmological surveys, much tighter constraints on cosmological parameters can be obtained than those derived from purely $2$-point measurements.

\bsp	
\label{lastpage}
\end{document}